\documentclass[journal]{IEEEtran}
\usepackage{lipsum}
\usepackage[utf8]{inputenc}
\usepackage{amsmath,amsfonts,amssymb,bm}
\usepackage{graphicx}
\usepackage{subcaption}
\usepackage{xcolor}
\usepackage{mathtools}
\usepackage{comment}
\usepackage{textcomp}

\usepackage{multirow}
\usepackage{epstopdf}

\newcommand{\bi}{\begin{itemize}}
\newcommand{\ei}{\end{itemize}}

\newcommand{\be}{\begin{enumerate}}
\newcommand{\ee}{\end{enumerate}}
\newcommand{\bd}{\begin{description}}
\newcommand{\ed}{\end{description}}
\newcommand{\bc}{\begin{center}}
\newcommand{\ec}{\end{center}}
\newcommand{\bt}{\begin{tabbing}}
\newcommand{\et}{\end{tabbing}}
\newcommand{\bfig}{\begin{figure}}
\newcommand{\efig}{\end{figure}}
\newcommand{\beq}{\begin{equation}}
\newcommand{\beqarr}{\begin{eqnarray}}
\newcommand{\beqarrn}{\begin{eqnarray*}}
\newcommand{\eeq}{\end{equation}}
\newcommand{\eeqarr}{\end{eqnarray}}
\newcommand{\eeqarrn}{\end{eqnarray*}}
\newcommand{\bflr}{\begin{flushright}\vspace{-0.2in}}
\newcommand{\eflr}{\end{flushright}}
\newcommand{\bsub}{\begin{subequations}}
\newcommand{\esub}{\end{subequations}}
\newcommand{\barr}{\begin{array}}
\newcommand{\earr}{\end{array}}
\newcommand{\nn}{\nonumber}

\def\undb#1{\mbox{\bf{#1}}}

\def\BibTeX{{\rm B\kern-.05em{\sc i\kern-.025em b}\kern-.08em
		T\kern-.1667em\lower.7ex\hbox{E}\kern-.125emX}}

\begin{document}

\title{RIS-Assisted THz MIMO Wireless System in the Presence of Direct Link for CV-QKD with Limited Quantum Memory}
\author{Sushil Kumar and Soumya~P.~Dash,~\IEEEmembership{Senior Member,~IEEE}

\thanks{The authors are with the School of Electrical and Computer Sciences, Indian Institute of Technology Bhubaneswar, Argul, Khordha, 752050 India e-mail: (22sp06003@iitbbs.ac.in, spdash@iitbbs.ac.in).}
}
\maketitle

\begin{abstract}
A reconfigurable intelligent surface (RIS)-aided multiple-input multiple-output (MIMO) wireless communication system is considered in this paper wherein the transmitter, Alice modulates secret keys, by using a continuous variable quantum key distribution technique to be transmitted to the receiver, Bob, which employs homodyne detection for data decoding. The data is transmitted over two paths, namely a direct path between Alice and Bob and the wireless path between them via the RIS. Transmit and receive beamsplitters are employed in the system to transform the MIMO terahertz channels into parallel single-input single-output channels. Considering an eavesdropper, Eve, to attack all the three wireless channels in the system (i.e., the direct channel, the channel between Alice and RIS, and between the RIS and Bob) but having restricted quantum memory limiting it to store the ancilla modes from either of these three wireless channels, novel expressions for the secret key rate (SKR) of the system are derived. Numerical results are presented to demonstrate the dependency of the system's performance on various system parameters. It is observed that the RIS plays a key role in increasing the SKR of the system and the transmission distance, ensuring secure communications between Alice and Bob. The significance of employing RIS is observed specifically for the case when Eve measures the ancilla modes of the channel between the RIS and Bob. Furthermore, for all such measurement scenarios, optimal angles are obtained for the phase shifts of the RIS elements to maximize the SKR for various MIMO configurations and transmission distance between Alice and Bob.
\end{abstract}
\section{Introduction}
The advent of novel technologies for beyond fifth-generation (B5G) and sixth-generation (6G) wireless communication systems is driven by the requirements of achieving high data rates, ultra-low latency, and high energy and spectral efficiencies \cite{GiPoMe:cm20,9669056}. 
This has led to an increased interest in developing various physical layer techniques utilizing massive multiple-input multiple-output (MIMO) \cite{9825647, 9585108}, terahertz (THz) frequency \cite{9766110, 9546670}, integrated sensing and communications (ISAC) \cite{9982596, 10042240}, non-coherent communications \cite{10302402, 8437142, 9689998}, and reconfigurable intelligent surfaces (RIS) \cite{9247315, 10256045}.
Of these, RIS has shown an immense potential to revolutionize the wireless communication scenario by utilizing a vast array of intelligent passive components, such as meta-surfaces and meta-materials, that are optimized electronically through real-time interaction with electromagnetic waves to manipulate the physical wireless channel \cite{10255749, 9847080}. Thus, a RIS can strategically direct the beams toward the designated users, thereby ensuring reliable communication with increased data rates, improved signal quality, extended coverage, and minimized interference \cite{10385147}. Moreover, implementing RIS technology enables robust communication links for systems even with manageable non-line-of-sight (NLoS) paths. This has led to the usage of RIS to improve the performance of MIMO systems, ISAC systems, index-modulated systems, and non-terrestrial networks \cite{10054402, 9698029, 9802114, 10400440, 10388479, 10287142}.

Another crucial aspect and requirement for the next-generation communication systems is the improvement of security and privacy of the transmitted data. The traditional higher-layer encryption techniques based on the Rivest-Shamir-Adleman (RSA) algorithm have been proven to be inefficient and easily decodable by utilizing Shor's algorithm owing to the rapid technological advancements in quantum computing \cite{weedbrook2012gaussian, manzalini2020quantum}. Moreover, the development of quantum computers has resulted in the usage of algorithms like Diffie and Hellman for physical layer encryption to be inefficient due to their capacity to yield solutions to computationally hard discrete logarithmic problems in a short time \cite{diffie1976new}. To overcome these challenges and enhance data privacy, quantum key distribution (QKD), a technique leveraging the underlying principles of quantum superposition and entanglement, has been proposed in the literature to offer unconditionally secure communication \cite{8865103, 9845436, 9552894}.

QKD ensures secure transmission of secret keys between two authenticated users, namely Alice and Bob, even in the presence of a potential eavesdropper, Eve. The literature on QKD classifies the technique into two categories, i.e., discrete variable QKD (DV-QKD) and continuous variable QKD (CV-QKD) \cite{8865103, cv_dv_qkd_Scarani_2009}. 
The DV-QKD approach relies on sources and detectors specifically designed for single photons and encrypts confidential key information by utilizing the polarisation or phase of a single photon. The secure key produced by the DV-QKD approach is guaranteed by the no-cloning theorem of quantum physics, preventing the possibility of making perfect replicas of non-orthogonal quantum states without introducing detectable noise \cite{9102386, 9799745}. On the contrary, secret key is encoded using the quadratures of continuous variable Gaussian quantum states in the CV-QKD technique, and its security is ensured by Heisenberg's uncertainty principle \cite{weedbrook2012gaussian, 8732438}. Owing to higher hardware compatibility with classical communication systems and the ability to offer superior key rates, performance in noisy environments, and ease of implementation at higher frequency ranges, CV-QKD becomes a natural choice for wireless systems over DV-QKD \cite{lodewyck2007quantum, cvqkd_ppf_2024}.

A majority of the wireless systems employing QKD focus on applications requiring point-to-point communications, such as satellite-to-earth channels, channels between two satellites, inter-building channels, and free-space maritime channels \cite{pirandola2021limits, pan2020secretOptics, pirandola2021satellite, gariano2017engineering, gariano2018trade}, which are traditionally achieved by employing free-space optical communications (FSO). However, it has been studied that THz communication is often preferred over FSO due to its better resilience to weather, NLoS capabilities, easier deployment (with less strict alignment), and ability to handle high data rates over short distances \cite{mmWaveThzMIMO_pirandola_10104156}.
This has led to several studies reported in the literature considering the usage of the THz band for secure communications using the CV-QKD technique for terrestrial and inter-satellite wireless systems \cite{ottaviani2020terahertz,inter_satellite_QKD_thz_Wang_2019, satellite2ground_CVQKD_10415457}. However, these studies have reported achieving a low secret key rate (SKR) and low transmission distance of the secret keys due to several factors resulting in channel degradation in the considered frequency spectrum. These limitations have been shown to be overcome to some extent by the integration of MIMO with CV-QKD \cite{2_ref_paper, neel_spd_channel_estimation_skr}. On another front, RIS has also been integrated with QKD, wherein the authors in \cite{Qkd_with_ris_Sayeed:22} introduce the idea of reduction of reflection loss with the use of RIS, and the authors in \cite{ neel_IRS_assisted_nlosqkd_10289124} employ RIS to aid QKD-based quantum communication in a FSO system. However, to the best of our knowledge, no reported study has analyzed the advantage of integrating RIS and MIMO techniques for a CV-QKD-aided wireless communication system operating in the THz frequency band, thus bringing out the full advantage of integrating these next-generation wireless technologies.

Inspired by this research gap, we consider a  RIS-assisted MIMO CV-QKD wireless communication system where the RIS assists the secret key transmission between Alice and Bob apart from the direct communication channel available between them. An eavesdropper, Eve, is considered to have the ability to attack all the available channels between Alice and Bob to decrypt the secret key encoded by the CV-QKD technique. In summary, the major contributions of the paper are as follows:
\begin{itemize}
\item A RIS-assisted $N_{R_X} \times N_{T_X}$ MIMO CV-QKD system model is proposed wherein singular-value decomposition (SVD)-based beamsplitters at the transceiver pair are employed to transform the THz MIMO system into multiple parallel single-input single-output (SISO) channels.
\item Eve is considered to employ a collective Gaussian entanglement attack at all the wireless channels available between Alice and Bob, and Bob utilizes homodyne detection for secret key decoding.
\item Considering a limited mode storage capacity of Eve constraining her in storing the ancilla modes from one of the wireless channels, novel expressions for the SKR for the proposed system are derived using the measurements at Bob and the stored ancilla modes of Eve.
\item Extensive numerical simulations are carried out to corroborate the analysis carried out in the paper. The effect of the increase in the number of RIS elements on the SKR and the transmission distance between Alice and Bob, the dependency of the measurement choice of the channel on the improvement of such performance, and the role of the phases of the RIS on the SKR of the system are demonstrated via numerical results.
\end{itemize}
A comparative study reveals that employing RIS for the THz MIMO CV-QKD system significantly improves the SKR and the transmission distance between Alice and Bob. Furthermore, the impact of the RIS becomes sacrosanct for the case when Eve utilizes the Ancilla mode measurements for the RIS-Bob channel for secret key decryption. Furthermore, optimal RIS phase shifts are observed, which enhance the SKR and transmission distance for the considered wireless communication system.

The rest of the paper is organized as follows. The system model of the RIS-assisted MIMO CV-QKD system, the transmission of the secret keys between Alice and Bob, and the attack carried out by Eve are presented in Section II. The closed-form expressions for the SKR of the system owing to Eve storing the ancillary modes of various possible channels are derived in Section III. Section IV presents the numerical results, followed by concluding remarks in Section V.

\textit{Notation;} Bold $(\textbf{A})$ letters stand for matrices. The $\textbf{A}^\dagger$ symbol stands for the conjugate transpose of $\textbf{A}$, and the $\textbf{A}^T$ symbol stands for the transpose. The notations $\boldsymbol{1}_{M\times N}$ and $\boldsymbol{0}_{M\times N} \in \mathbb C^{M\times N}$ represent a matrix consisting of all ones and all zeros, respectively, and $\jmath=\sqrt{-1}$. $\textbf{I}_M$ represents a $M \times M$ identity matrix, and $\text{diag}(\boldsymbol{a})$ with $\boldsymbol{a} \in \mathbb C^M$ returns a $M \times M$ diagonal matrix with the elements of $\boldsymbol{a}$ on its principal diagonal. $a^*$ represents the conjugate of $a$. The notation $\langle X \cdot Y\rangle$ represents the quantum correlation between X and Y. The notation $\mathcal{N} (\boldsymbol{\mu},\boldsymbol{\sigma^2})$ represents the real multivariate Gaussian distribution, where $\boldsymbol{\mu}$ is the mean vector and $\boldsymbol{\sigma^2}$ is the covariance matrix, $|\cdot|$ denotes the magnitude operator, and the operator $\text{eig} (\cdot)$ computes the eigenvalues.
\section{System Model}
\subsection{Channel Model}
\begin{figure}[t!]
    \centering
    \includegraphics[width=8cm, height=5cm]{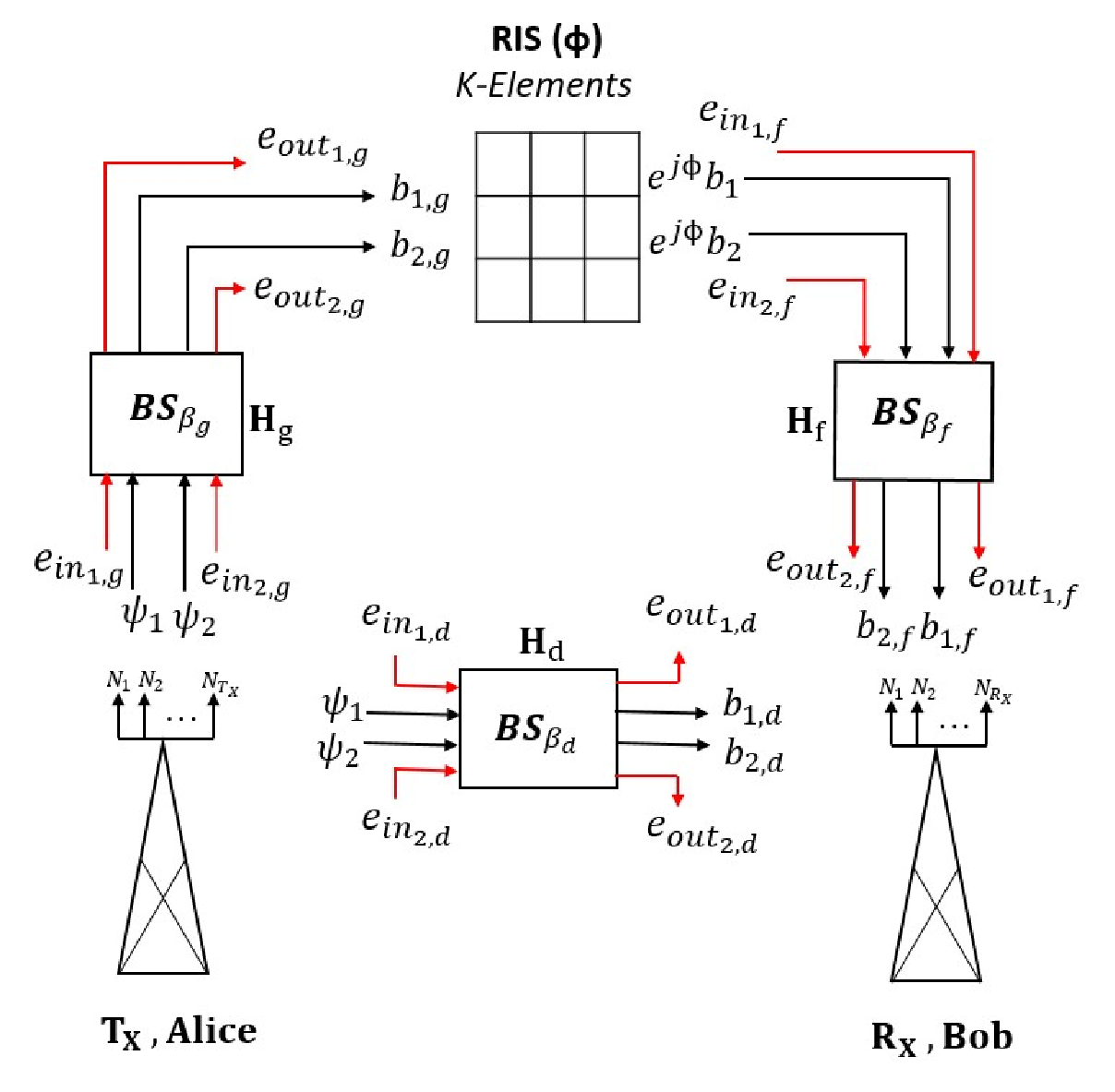}
    \caption{Model of the RIS-assisted MIMO CV-QKD wireless communication system.}
    \label{f1}
\end{figure}

We consider a RIS-assisted MIMO CV-QKD system as shown in Fig. \ref{f1}, where the transmitter, termed Alice, and the receiver, termed Bob, comprise of $N_{T_X}$ and $N_{R_X}$ antennas, respectively.
The RIS consists of $K$ passive reflecting elements, which are configured to adjust the phases of the incident and the transmitted electromagnetic signals. The communication is also aided by the direct LoS path present between the transceiver pair. Thus, the effective MIMO channel between Alice and Bob, denoted by $\textbf{H}\in \mathbb{C}^{N_{R_X}\times N_{T_X}}$, is obtained as
\beq
\textbf{H} = \textbf{H}_{d} + \textbf{H}_{f} \mathbf{\Phi}_{\text{RIS}} \textbf{H}_{g} \, ,
\label{eq1}
\eeq
where $\mathbf{\Phi}_{\text{RIS}} = \text{diag} \left(e^{\jmath \phi_{1}},...,e^{\jmath \phi_{K}}\right)$ and $\phi_k$ is the phase shift introduced by the $k$-th element of the RIS. Furthermore, $\textbf{H}_{d}\in {\mathbb{C}}^{N_{R_X} \times N_{T_X}}$, $\textbf{H}_{g} \in \mathbb{C}^{K\times N_{T_X}}$, and $\textbf{H}_{f}\in \mathbb{C}^{N_{R_X} \times K}$, are the direct LoS channel matrix between the transceiver pair, the channel matrix between Alice and RIS, and the channel matrix between RIS and Bob, respectively, which are expressed as
\beqarr
&& \! \! \! \! \! \! \! \! \! \!
\! \! \! \! \! \! \! \! \! \! \textbf{H}_{d} =
\sum_{\ell=1}^L \sqrt{\delta_{d,\ell}}
e^{\jmath 2 \pi f_c \tau_{d,\ell}} 
\textbf{h}_{N_{R_X}}\left( \theta_\ell^{R_X} \right)\textbf{h}^\dagger_{N_{T_X}}
\left(\theta_{\ell}^{T_X} \right) \, , \nn \\
&& \! \! \! \! \! \! \! \! \! \!
\! \! \! \! \! \! \! \! \! \! \textbf{H}_{g} =
\sum_{m=1}^M \sqrt{\delta_{g,m}}
e^{\jmath 2 \pi f_c \tau_{g,m}} \textbf{h}_{\text{RIS}} 
\left(\varphi,\theta_m^{\text{RIS}}\right)
\textbf{h}^\dagger_{N_{T_X}}
\left( \theta_m^{T_X} \right) \, , \nn \\
&& \! \! \! \! \! \! \! \! \! \!
\! \! \! \! \! \! \! \! \! \! \textbf{H}_{f} =
\sum_{n=1}^N \sqrt{\delta_{f,n}}
e^{\jmath 2 \pi f_c \tau_{f,n}}
\textbf{h}_{N_{R_X}} \left( \theta_n^{R_X} \right) 
\textbf{h}^\dagger_{\text{RIS}}
\left( \varphi,\theta_n^{\text{RIS}} \right) \, .
\label{eq2}
\eeqarr
Here $L, M$, and $N$ denote the number of multipaths in the wireless channels $\textbf{H}_d$, $\textbf{H}_g$, and $\textbf{H}_f$, respectively. Further, $f_c$ is the carrier frequency, and $\tau_{d}$, $\tau_{g}$, $\tau_{f}$ and $\delta_d$, $\delta_{g}$, $\delta_{f}$ are the propagation delays and path losses corresponding to $\textbf{H}_d$, $\textbf{H}_g$, and $\textbf{H}_f$, respectively. Moreover, $\theta_{\ell}^{T_X}$, $\theta_m^{T_X}$, and $\theta_n^{\text{RIS}}$ are the angles of departure (AoD) and $\theta_{\ell}^{R_X}$, $\theta_m^{\text{RIS}}$, and $\theta_n^{R_X}$ are the angles of arrival (AoA) of the $\ell$-th, $m$-th, and $n$-th multipath from the transmitter's and the receiver's uniform linear arrays (ULAs), respectively. The antenna elements in both ULAs are uniformly placed in a single dimension such that the inter-element spacing is maintained at $d_a$. This results in the array response vector in (\ref{eq2}), i.e., $\undb{h}_{N_{T_X}} \left( \text{or } \undb{h}_{N_{R_X}} \right)$, to be given as
\beq
\undb{h}_{N} \left( \theta \right) = \frac{1}{\sqrt{N}} 
\left[1, e^{\jmath \frac{2\pi}{\lambda_c}
d_a \sin \theta} , \cdots ,
e^{\jmath \frac{2\pi}{\lambda_c}
d_a \left( N-1 \right) \sin \theta} \right]^T ,
\label{eq3}
\eeq
where $N \in \{N_{T_X},N_{R_X}\}$ and $\lambda_c = c/f_c$, with $c$ being the speed of electromagnetic signals. Furthermore, $\textbf{h}_{\text{RIS}}\in \mathbb C^{K\times 1}$ in (\ref{eq2}) is the response vector of the passive ULA of the RIS \cite{path_los_RIS_Ellingson_2021}, \cite{ris_assisted_mmwave_mimo_9918631} and is given as
\beqarr
&& \! \! \! \! \! \! \! \! \!
\! \! \! \! \! \! \! \! \! \!
\textbf{h}_{\text{RIS}} 
\left( \varphi, \theta_{i}^{\text{RIS}} \right)
= \frac{1}{\sqrt{K}} \left[e^{\jmath\frac{2\pi}{\lambda_c}
\left(\vartheta_X^{\varphi, \theta_{i}^{\text{RIS}}} 
+ \vartheta_Y^{\varphi, \theta_{i}^{\text{RIS}}} \right)}, \cdots \right. \nn \\
&& \qquad \left. \cdots ,
e^{\jmath\frac{2\pi}{\lambda_c}
\left( \left( K_X-1 \right) \vartheta_X^{\varphi, \theta_{i}^{\text{RIS}}}
+ \left( K_Y - 1 \right) \vartheta_Y^{\varphi, \theta_{i}^{\text{RIS}}} \right)} \right] ,
\label{eq4}
\eeqarr
where 
\beqarr 
\vartheta_X^{\varphi, \theta_{i}^{\text{RIS}}}
\! \! \! \! &=& \! \! \! \! d_X \cos\left(\varphi\right)
\sin\left(\theta_{i}^{\text{RIS}} \right) ,\nn\\
\vartheta_Y^{\varphi, \theta_{i}^{\text{RIS}}}
\! \! \! \! &=& \! \! \! \! d_Y \sin\left(\varphi\right)
\sin \left(\theta_i^{\text{RIS}} \right) \, , i \in \left\{m , n \right\} \, .
\label{eq5}
\eeqarr
Here the elements of the RIS are considered to be arranged along a 2-dimensional structure \cite{ris_review_2023, ris_channel} with $K_X$ and $K_Y$ reflecting elements along the horizontal and the vertical axes, respectively, implying that $K_X K_Y=K$ and the separation between the elements in the corresponding axes is denoted by $d_X$ and $d_Y$. Furthermore, the path losses in (\ref{eq2}) are expressed as
\beq
\delta_{j,i} \! = \!
\begin{cases} \! \! 
    \left( \frac{\lambda_c}{4\pi d_{j,i}} \right)^2 
    \! \! G_{T_X} G_{R_X} 10^{-0.1 \rho d_{j,i}} \, , j \in \left\{ d, g, f \right\} , i=1 \\
    \! \varsigma \xi_i \left( \frac{\lambda_c}
    {4\pi d_{j,i}} \right)^2 \! \! G_{T_X} G_{R_X}
    10^{-0.1 \rho d_{j,i}}
    \! , i \in \left\{ \ell, m, n \neq 1 \right\} \! ,
\end{cases}
\label{eq6}
\eeq
where $d_{j,i}$ is the smallest path length,  $\xi_i$ is the Fresnel reflection coefficient of the $i$-th multipath component, $\varsigma$ is the Rayleigh roughness factor, and $\rho$ (in dB/km) is the atmospheric absorption loss. Moreover, Alice's and Bob's ULAs gains are $G_{T_X} = N_{T_X}G_a$ and $G_{R_X} = N_{R_X}G_a$.
respectively, where $G_a$ is the gain of each component of the transmitter and receiver antennas and $G_{T_X}$ (or $G_{R_X}) = K$ when the data symbols are transmitter from (or to) the RIS, respectively.
\subsection{Secret Key Generation, Eve's Attack, and Bob's Measurement}
In the scenario of this MIMO wireless communication system, Alice aims to transmit a secure quantum key to Bob. Towards this end, Alice employs a Gaussian modulated CV-QKD scheme to create two independent zero-mean Gaussian random vectors as $\undb{X}_{\text{Alice}}, \undb{P}_{\text{Alice}} \sim {\mathcal{N}} \left( \mathbf{0}_{N_{T_X}}, V_s \undb{I}_{N_{T_X}} \right)$, which correspond to the position and the momentum quadratures, respectively. 
Furthermore, Alice transmits the coherent states \cite{2_ref_paper, QC_book} $\psi_t = X_{\text{Alice},t} + \jmath P_{\text{Alice},t}$ for $t = 1, \ldots, N_{T_X}$ from the $N_{T_X}$ antennas. These coherent states are passed through the beam-splitters $\boldsymbol{{BS_\beta}}_d$ and $\boldsymbol{{BS_\beta}}_g$ of channels $\textbf{H}_{d}$ and $\textbf{H}_{g}$, respectively, and the reflected signals from the RIS are passed through the beam-splitter $\boldsymbol{{BS_\beta}}_f$ of channel $\textbf{H}_{f}$, as shown in Fig. \ref{f1}. The eavesdropper, termed Eve, attacks all the wireless channels to decrypt the transmitted secret key by creating two-mode squeezed vacuum states (TMSV), also known as entangled Einstein-Poldolsky-Rosen (EPR) pairs, as $\{e_{in_1},e_{qm_1}\} \ldots \{e_{in_{N_{T_X}}},e_{qm_{N_{T_X}}}\}$ for each coherent state transmitted by Alice \cite{2_ref_paper, ref-7}. Further, $e_{in_1},\ldots e_{in_{N_{T_X}}}$ are utilized by the Eve to eavesdrop into the transmitted secret key and $e_{qm_1}, \ldots, e_{qm_{N_{T_X}}}$ are the ancillary modes stored in the quantum memory of Eve.  

\begin{figure}[!t]
    \centering
    \includegraphics[height=1.4in,width=3.2in]{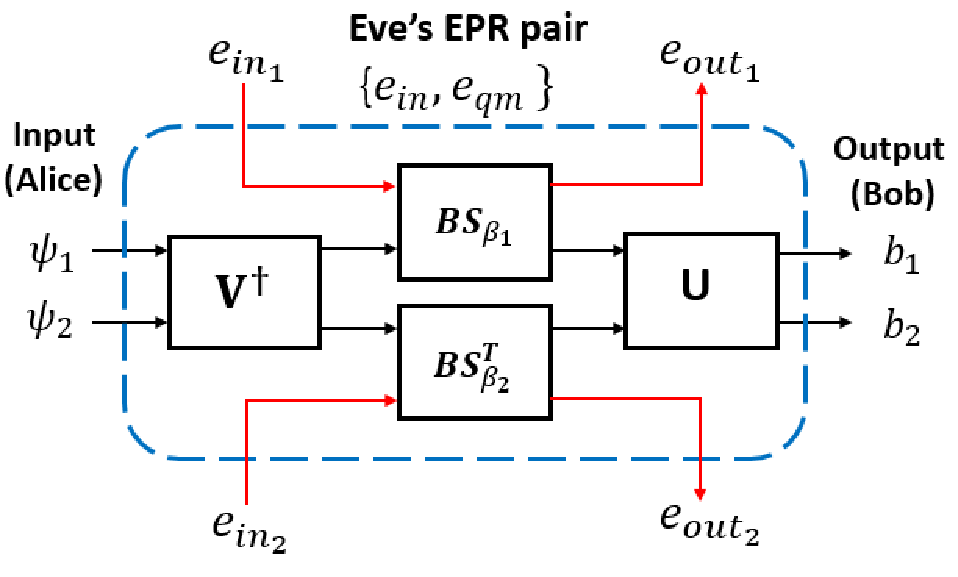}
    \caption{A depiction of four beam-splitter models representing a $2 \times 2$ MIMO configuration channel.}
    \label{f2}
\end{figure}

For a better understanding of secret-key communication in the presence of Eve in the RIS-assisted MIMO system under consideration, we consider an example of a $2 \times 2$ MIMO configuration. Such a system can thus be
modeled by four two-port balanced beamsplitters, as shown in Fig. \ref{f2}. The first input beamsplitter, denoted as $\textbf{V}^\dagger$, combines the two coherent states $\{\psi_1, \psi_2\}$ sent by Alice and produces two output modes. The next set of beamsplitters, $\boldsymbol{{BS_\beta}}_1$ and $\boldsymbol{{BS_\beta}}_2$, combine the output of the first beamsplitter with Eve's modes $e_{in_1}$ and $e_{in_2}$, respectively. Furthermore, the output beamsplitter $\textbf{U}$ gives the output $\{b_1,b_2\}$ which can be expressed as \cite{beamsplitter, jantti2017multiantenna}
\beq
\begin{bmatrix} b_{1} \\ b_{2} \end{bmatrix}
= \begin{bmatrix} \sqrt{\beta} & \sqrt{1-\beta} \\
- \sqrt{1-\beta} & \sqrt{\beta} \end{bmatrix}
\begin{bmatrix} \psi_1 \\ \psi_2 \end{bmatrix} \, ,
\label{eq7}
\eeq
where $\beta$ denotes the overall transmissivity of the beamsplitter system.

Extending this formulation to the RIS-assisted $N_{T_X} \times N_{R_X}$ MIMO system in the presence of the direct channel link, the $N_{R_X} \times 1$ received signal vector mode at Bob, denoted by $\undb{b}$, is obtained as
\beqarr
\textbf{b} \! \! \! \! &\stackrel{(a)}{=}& \! \! \! \! \left(\textbf{H}_d + \textbf{H}_f 
\mathbf{\Phi}_{\text{RIS}} \textbf{H}_g\right) \Psi \nn \\
&& + \left(\textbf{U}_d\textbf{S}_d
+ \textbf{H}_f \mathbf{\Phi}_{\text{RIS}} \textbf{U}_g \textbf{S}_g 
+\textbf{U}_f\textbf{S}_f\right) \textbf{e} \nn \\
&\stackrel{(b)}{=}& \! \! \! \!
\left(\textbf{U}_d \textbf{D}_d \textbf{V}_d^\dagger
+ \textbf{U}_f \textbf{D}_f \textbf{V}_f^\dagger \mathbf{\Phi}_{\text{RIS}} 
\textbf{U}_g \textbf{D}_g \textbf{V}_g^\dagger\right) \Psi \nn \\
&& + \left(\textbf{U}_d\textbf{S}_d+\textbf{U}_f\textbf{D}_f\textbf{V}_f^\dagger 
\mathbf{\Phi}_{\text{RIS}} \textbf{U}_g\textbf{S}_g 
+\textbf{U}_f\textbf{S}_f\right) \textbf{e} \, ,
\label{eq8}
\eeqarr
where step $(a)$ in (\ref{eq8}) is obtained owing to the transmission of the coherent state vector $\Psi = \left[ \psi_1, \ldots, \psi_{N_{T_X}} \right]^T$ by Alice and the injection of eavesdropping Gaussian mode vector $\undb{e} = \left[ e_{in_1},\ldots, e_{in_{N_{T_X}}} \right]^T$ by Eve at the corresponding beamsplitters in each channel. Moreover, step $(b)$ occurs by expressing the channel matrices in terms of their singular value decompositions (SVDs) as $\textbf{H}_j = \textbf{U}_j\textbf{D}_j\textbf{V}_j^\dagger$, $\forall j \in \{d,g,f\}$ where $\undb{U}_j$s $\in \mathbb C^{R_X\times R_X}$ and $\undb{V}_j$s $\in \mathbb C^{T_X\times T_X}$ are unitary matrices with $R_X=N_{R_X}$ for $\undb{H}_d,\undb{H}_f$ and $R_X=K$ for $\undb{H}_g$, and $T_X=N_{T_X}$ for $\undb{H}_d,\undb{H}_g$ and $T_X=K$ for $\undb{H}_f$. Further, $\undb{D}_j$s are the matrices containing the singular values for the corresponding channels and are given as
\beq
\textbf{D}_j =
\begin{bmatrix}
    \text{diag}\left(\sqrt{\beta_{j_1}},\ldots ,\sqrt{\beta_{j_r}}\right)&\boldsymbol{0}_{r \times\left(T_X-r\right)}\\
    \boldsymbol{0}_{\left(R_X -r\right) \times r}&\boldsymbol{0}_{\left(R_X -r\right) \times (T_X-r)}
\end{bmatrix} \, ,
\label{eq9}
\eeq 
where $r$ and $\sqrt{\beta_{j_1}},...,\sqrt{\beta_{j_r}}$ are the rank and the $r$ non-zero singular values of the $j$-th channel matrix. Additionally, the diagonal matrix $\textbf{S}_j$ is given as
\beq
\textbf{S}_j =
\text{diag}\left(\sqrt{1-\beta_{j_1}}, \ldots, \sqrt{1-\beta_{j_r}}, \underbrace{1,\ldots,1}_{\left(N_j-r\right) \text{ times}} \right) \, ,
\label{eq10}
\eeq 
where
$$N_j \in \left\{\min\left(N_{T_X} , N_{R_X}\right), \min\left(N_{T_X}, K\right), \min\left(K, N_{R_X}\right)\right\}$$
for $j\in \{d,g,f\}$, respectively.

It is to be noted that to extract the secret key information from the communication channel, Eve uses a collective Gaussian entanglement attack on all wireless channels, which is considered the most powerful attack against the Gaussian CV-QKD protocols \cite{2_ref_paper}, \cite{weedbrook2012gaussian}. To illustrate its effect we again consider the example of the $2 \times 2$ MIMO case wherein the first modes of Eve's created EPR pair $e_{in_1}$ and $e_{in_2}$ are mixed with the incoming coherent states of Alice with the help of the beamsplitters $\boldsymbol{{BS_\beta}}_1$ and $\boldsymbol{{BS_\beta}}_2$, respectively, and the second modes, $e_{qm_1}$ and $e_{qm_2}$, are stored in her quantum memory. Further, the outputs from the beamsplitters, namely $e_{out_1}$ and $e_{out_2}$ are also stored in Eve's quantum memory and the other two outputs are passed through the final beamsplitter $\undb{U}$. Generalizing it for the MIMO system, we consider a scenario of Eve having a restricted memory \cite{neel_restricted_qkd, pan2020secret, pan2021geometrical}. This scenario is also advantageous for quantum systems due to the fact that the stored ancilla modes become more prone to errors with the increase in quantum memory \cite{quantum_memory2023acharya}. 
Thus, Eve stores one of the output state vectors at the same time: $\undb{e}_{out_{d}}$, $\undb{e}_{out_{g}}$, and $\undb{e}_{out_{f}}$ generated via the channels $\textbf{H}_{d}$, $\textbf{H}_{g}$, and $\textbf{H}_{f}$, respectively. This results in the ancilla modes of Eve to be generated from her quantum memory containing the output modes and stored states as $\{\undb{e}_{out_j}, \undb{e}_{qm}\}$, for $j \in \{d, g,f\}$.

We consider the scenario where Alice and Bob have the perfect knowledge of the fading channels \cite{2_ref_paper,neel_restricted_qkd}. Following this, Alice employs transmit beamforming matrices $\textbf{V}_d$ and $\textbf{V}_g$ for the direct path and RIS path, respectively, and Bob employs combiners as $\textbf{U}_d^\dagger$ and $\textbf{U}_f^\dagger$ for the received data from the direct path and RIS path, respectively. This results in the output vector at Bob to be given as
\beqarr
\textbf{b} \!\!\!\!&=&\!\!\!\! \left(\textbf{U}_d^\dagger\textbf{H}_d \textbf{V}_d 
+ \textbf{U}_f^\dagger\textbf{H}_f \mathbf{\Phi}_{\text{RIS}} \textbf{H}_g \textbf{V}_g\right) \Psi \nn \\
&+& \!\!\!\! \left(\textbf{U}_d^\dagger\textbf{U}_d\textbf{S}_d
+\textbf{U}_f^\dagger\textbf{H}_f \mathbf{\Phi}_{\text{RIS}} 
\textbf{U}_g\textbf{S}_g 
+ \textbf{U}_f^\dagger \textbf{U}_f \textbf{S}_f \right) \textbf{e} .
\label{eq11}
\eeqarr 
Utilizing SVD of the channel matrices $\textbf{H}_d, \textbf{H}_g$, and $\textbf{H}_f$ modifies the MIMO channel into equivalent $r$ parallel SISO channels, and thus, (\ref{eq11}) can be re-written as
\beqarr
&& \! \! \! \! \! \! \! \! \! \! \! \! \!
b_i = \underbrace{\left(\sqrt{\beta_{f_i} \beta_{g_i}}
e^{\jmath \phi_i} \right) \!\psi_i
\!+\!\left(\!\!\sqrt{1\!-\!\beta_{f_i}}
\!+\!\sqrt{\beta_{f_i}\!\left(\!1\!-\!\beta_{g_i}\right)}e^{\jmath \phi_i} \!\!\right) \!e_{in_i}}_{b_{{\text{RIS}}_i}} \nn \\
&& + \underbrace{\left(\sqrt{\beta_{d_i}} \psi_i
+ \sqrt{1-\beta_{d_i}}e_{in_i}\right)}_{b_{d_i}} \, , i=1,\ldots, r \, ,
\label{eq12}
\eeqarr
where $b_{d_i}$s and $b_{\text{RIS}_i}$s are the received signals at Bob via the direct path and the RIS paths, respectively, and $\beta_{j_i}$ denotes the transmissivity of the $i$-th parallel channel in the $j$-th wireless channel.

It is to be recalled that Eve attacks at all the channels with the same generated EPR pair mode $\{\undb{e}_{in},\undb{e}_{qm}\}$. Thus, the input-output expression for Eve's output modes corresponding to channels $\textbf{H}_j$s for $j \in \{ d,g,f\}$, are given by
\beqarr
e_{out_{g_i}} \! \! \! \! &=& \! \! \! \!
-\sqrt{1-\beta_{g_i}} \psi_i + \sqrt{\beta_{g_i}}e_{in_i} \, , \nn \\
e_{out_{f_i}} \! \! \! \! &=& \! \! \! \!
-\sqrt{\left( 1-\beta_{f_i} \right) \beta_{g_i}} e^{\jmath \phi_i} \psi_i \nn \\
&&
+ \underbrace{\left(\sqrt{\beta_{f_i}} - \sqrt{\left( 1 - \beta_{g_i} \right) \left( 1-\beta_{f_i} \right)} e^{\jmath \phi_i} \right)}_{\tilde{\beta}_{f_i}} e_{in_i} \, , \nn \\
e_{out_{d_i}} \! \! \! \! &=& \! \! \! \!
-\sqrt{1-\beta_{d_i}} \psi_i + \sqrt{\beta_{d_i}}e_{in_i} \, , i=1,\ldots,r \, .
\label{eq13}
\eeqarr

Following the transmission of the secret key and the creation of the ancilla modes, Bob proceeds to measure the quadratures transmitted by Alice. There are two types of measurements that Bob can perform, namely 1) homodyne, in which the measurement is performed on one of the two quadratures randomly, and 2) heterodyne, where measurement is performed on both quadratures simultaneously \cite{QC_book, collective_GA_Pirandola_2008}. The study conducted in \cite{neel_spd_channel_estimation_skr} demonstrates that both homodyne and heterodyne measurements yield similar results owing to an increased detector noise in heterodyne measurements. Therefore, in this paper, we consider that Bob performs homodyne measurements. This results in the corresponding measurements for the $\undb{H}_j$-th sub-channel for $j \in \{d,g,f\}$ as
\bsub
\beq
\hat{Q}_{b_i} = \hat{Q}_{b_{d_i}} + \hat{Q}_{b_{{\text{RIS}}_i}} \, , i=1,\ldots,r \, ,
\label{eq14a}
\eeq
where
\beqarr
&& \! \! \! \! \! \! \! \! \! \! \! \! \! \! \! \! \! \! \! \!
\hat{Q}_{b_{d_i}} = 
\sqrt{\beta_{d_i}} \hat{Q}_{a_i}+ \sqrt{1-\beta_{d_i}} \hat{Q}_{e_{in_i}}, \nn \\
&& \! \! \! \! \! \! \! \! \! \! \! \! \! \! \! \! \! \! \! \!
\hat{Q}_{b_{{\text{RIS}}_i}}
= \sqrt{\beta_{f_i} \beta_{g_i}} e^{\jmath \phi_i} \hat{Q}_{a_i} \nn \\
&& 
+ \left(\sqrt{1-\beta_{f_i}} + \sqrt{\beta_{f_i} \!\left( 1\!-\!\beta_{g_i}\right)}e^{\jmath \phi_i}\right) \! \hat{Q}_{e_{in_i}} ,
\label{eq14b}
\eeqarr
\esub
and the measurements of the ancillary mode of Eve to be given as
\beqarr
\hat{Q}_{e_{out_{g_i}}} \! \! \! \! &=& \! \! \! \! -\sqrt{\left(1-\beta_{g_i}\right)} \hat{Q}_{a_i} + \sqrt{\beta_{g_i}} \hat{Q}_{e_{in_i}} \, , \nn \\
\hat{Q}_{e_{out_{f_i}}} \! \! \! \! &=& \! \! \! \!
-\sqrt{\left(1-\beta_{f_i}\right) \beta_{g_i}} e^{\jmath \phi_i} \hat{Q}_{a_i} \nn \\
&&
\!\!\!+ \underbrace{\left(\sqrt{\beta_{f_i}} - \sqrt{\left( 1 - \beta_{g_i} \right) \left( 1-\beta_{f_i} \right)} e^{\jmath \phi_i} \right)}_{\tilde{\beta}_{f_i}}
\hat{Q}_{e_{in_i}} \, , \nn \\
\hat{Q}_{e_{out_{d_i}}} \! \! \! \! &=& \! \! \! \! -\sqrt{\left(1-\beta_{d_i}\right)} \hat{Q}_{a_i} + \sqrt{\beta_{d_i}} \hat{Q}_{e_{in_i}} \, , i = 1, \ldots, r , \nn \\
\label{eq15}
\eeqarr
where $\hat{Q}_{a}$, $\hat{Q}_{b}$, and $\hat{Q}_{e}$ represent one of the two quadratures (namely position and momentum) at Alice's, Bob's, and Eve's ends, respectively.
Let us consider the variance of Alice's transmitted mode and the variance of both the modes in Eve's EPR pair as $V_a \left(= V_s+V_o \right)$ and $V_e$, respectively, where $V_o$ is the variance of the thermal mode in the quantum channel. Thus, from (\ref{eq14b}), the variances of the received signals are obtained as
\bsub
\beqarr
V_{b_{d_i}} \! \! \! \! &=& \! \! \! \! \beta_{d_i} V_a + \left(1-\beta_{d_i}\right) V_e \, , \nn \\
V_{b_{{\text{RIS}}_i}} \! \! \! \! &=& \! \! \! \! \left(\alpha_i \alpha_i ^* \right)V_a
+ \left( \gamma_i \gamma_i^* \right) V_e \, , i=1,\ldots,r \, ,
\label{eq16a}
\eeqarr
where $V_{b_{d_i}}$ is Bob's variance from the $i$-th received quadrature via the direct LoS channel, $V_{b_{{\text{RIS}}_i}}$ is the variance from the RIS path, and we have
\beqarr
&& \! \! \! \! \! \! \! \! \! \! \! \! \! \! \! \! \! \! \! \!
\alpha_i = \sqrt{\beta_{g_i} \beta_{f_i}} e^{\jmath \phi_i} \, , \nn \\
&& \! \! \! \! \! \! \! \! \! \! \! \! \! \! \! \! \! \! \! \!
\gamma_i = \sqrt{1-\beta_{f_i}} 
+ \sqrt{\beta_{f_i}\left(1-\beta_{g_i}\right)}e^{\jmath \phi_i} \, , i = 1,\ldots, r \, .
\label{eq16b}
\eeqarr
\esub
Similarly, from (\ref{eq15}), the variances of the output modes of Eve are given as
\beqarr
V_{e_{{out}_{g_i}}} \! \! \! \! &=& \! \! \! \!
\left(1-\beta_{g_i}\right) V_a + \beta_{g_i} V_e \, , \nn \\
V_{e_{{out}_{f_i}}} \! \! \! \! &=& \! \! \! \!
\left(1-\beta_{f_i}\right) \beta_{g_i} V_a \nn \\
&+& \!\!\!\!\!\!\left(\!\!\left(1\!- \!\beta_{g_i}\right)\!+\!\beta_{g_i} \beta\!_{f_i}\!  - 2 \sqrt{\!\beta_{\!f_i} \!\left(\!1\! -\! \beta_{\!f_{i}}\! \right) \!\left(\!1\! -\! \beta_{g_i}\! \right)} \!\cos{\phi_i}\!\! \right) \!\!V_e \, , \nn \\
V_{e_{{out}_{d_i}}} \! \! \! \! &=& \! \! \! \!
\left(1-\beta_{d_i}\right) V_a + \beta_{d_i} V_e \, , \ i=1,\ldots,r \, .
\label{eq17}
\eeqarr
Furthermore, for each $i$-th parallel channel, the covariance matrix of Eve's ancillary correlated string $\{e_{out_{j,i}}, e_{qm_i}\}$, for $j \in \{d, g,f\}$, is denoted by $\undb{K}_{E_{j_i}}$ and can be expressed as
\beq
\undb{K}_{E_{j_i}} =
\begin{bmatrix}
    V_{e_{{out}_{j_i}}} \textbf{I}_2 & V_{ee_{j_i}} \mathbf{\Omega} \\
    V_{ee_{j_i}}^* \mathbf{\Omega}_r^T & V_{e} \textbf{I}_2
\end{bmatrix} \, , 
\label{eq18}
\eeq 
where
\bsub
\beq
V_{ee_{j_i}} =
\begin{cases} \! \! 
\sqrt{\beta_{j_i} \left(V_{e}^2-1\right)} \, , \
j \in \left\{ d, g \right\} , \\
\! \tilde{\beta}_{f_i} \sqrt{\left(V_{e}^2-1\right)} \, , \
j = f \, , \, i = 1, \ldots, r \, ,
\end{cases}
\label{eq19a}
\eeq
with
\beq
\mathbf{\Omega} = \begin{bmatrix} 0 & 1 \\ -1 & 0 \end{bmatrix} \, .
\label{eq19b}
\eeq
\esub
\section{Secret Key Rate}
It is to be recalled that Eve attacks all three wireless channels with her generated EPR pair to decrypt the secret key, and has a limited quantum memory size, implying that she can store only one of the three ancillary modes that contain the extracted information from Alice and Bob's ongoing conversation via data transmission, wherein the generated random vectors $\hat{\textbf{Q}}_{a}$ and $\hat{\textbf{Q}}_{b}$ are linked. Following the homodyne measurement, a reconciliation technique is employed by Bob to fix errors, which is typically of two types, namely, 1) direct reconciliation (DR) and 2) reverse reconciliation (RR) \cite{2_ref_paper,neel_spd_channel_estimation_skr, ref-7}. RR works better than DR because the RR protocol can achieve positive SKR for any channel transmittance $\beta_{j_i} \in [0,1]$. However, we require $\beta_{j_i} > 0.5$ to achieve a positive SKR using DR. Owing to this reason, we consider Bob to employ the RR protocol following the homodyne measurement, resulting in the expression of the effective SKR in the $i$-th channel to be given as 
\beq
\text{SKR}^{\text{RR}}_i = \left(\Delta I\right)_{i, \text{LoS}}
+ \left( \Delta I \right)_{i, \text{RIS}} \, , i = 1 \ldots, r \, ,
\label{eq20}
\eeq
where $\left(\Delta I\right)_{i,\text{LoS}}$ and $\left(\Delta I\right)_{i, \text{RIS}}$ denote the SKRs of the direct channel, $\textbf{H}_d$, and the channels through RIS, $\textbf{H}_g$ and $\textbf{H}_f$. It is to be noted that either for the LoS path or the RIS path, we have
\beq
\Delta I_i = I\left(Q_{a_i};Q_{b_i}\right) - I\left(Q_{b_i};E_i\right) \, , i=1,\ldots, r \, ,
\label{eq21}
\eeq
where $I\left(Q_{a_i}; Q_{b_i}\right)$ is the classical mutual information between Alice and Bob and $I\left(Q_{b_i}; E_i\right)$ is Eve's attainable information about Bob \cite{2_ref_paper}, \cite{QC_book}, \cite{ref-7} for the $i$-th channel. The classical Shannon mutual information between Alice and Bob in (\ref{eq21}) is given as
\beq
I\left(Q_{a_i};Q_{b_i}\right) = H(Q_{b_i}) - H\left(Q_{b_i}|Q_{a_i}\right) \, , i=1,\ldots, r \, ,
\label{eq22}
\eeq
where
\beq
H(Q_{b_i}) = \frac{1}{2} \log_2(V_{b_i}) \, , i=1\ldots, r \, ,
\label{eq23}
\eeq
and
\beq
H(Q_{b_i}|Q_{a_i}) = \frac{1}{2} \log_2(V_{b_i|a_i}) \, , i=1,\ldots, r \, ,
\label{eq24}
\eeq
\setcounter{equation}{31}
\begin{figure*}[t]
\beqarr
\lambda_{1,2} =
\begin{cases}
\sqrt{\frac{1}{2}
\left( V_{e_{out_{j_i}}}^2 + V_e^2 - 2 \beta_{j_i} \left( V_e^2 - 1 \right)
\pm \left( V_{e_{out_{j_i}}} - V_e \right)
\sqrt{\left( V_{e_{out_{j_i}}} + V_e \right)^2
- 4 \beta_{j_i} \left(V_e^2 - 1 \right)} \right) }
\, , \ j \in \left\{ d , g \right\} , \\
\sqrt{\frac{1}{2} \left( \begin{array}{cc}
& \hspace{-8cm}
V_{e_{out_{f_i}}}^2 + V_e^2
- 2 \Re \left\{ \tilde{\beta}_{j_i}^2 \right\} \left( V_e^2 - 1 \right) \\
& \! \! \! \! \! \! \! \! \!
\pm \sqrt{\left( \left( V_{e_{out_{j_i}}} + V_e \right)^2
- 4 \Re \left\{ \tilde{\beta}_{j_i} \right\}^2 \left(V_e^2 - 1 \right) \right)  
\left( \left( V_{e_{out_{j_i}}} - V_e \right)^2
+ 4 \Im \left\{ \tilde{\beta}_{j_i} \right\}^2 \left(V_e^2 - 1 \right) \right) }
\end{array} \right)} , j = f
\end{cases}
\label{eq32}
\eeqarr
\noindent\rule{\textwidth}{.5pt}
\vspace{-0.8cm}
\end{figure*}
with $V_{b_i}$ and $V_{b_i|a_i}$ denoting the variance of Bob's quadrature and the conditional variance of Bob's quadrature conditioned on the quadrature of Alice, respectively. Using (\ref{eq23}) and (\ref{eq24}) in (\ref{eq22}) results in the expression of the mutual information between Alice and Bob to be simplified as
\setcounter{equation}{24}
\beq
I\left(Q_{a_i};Q_{b_i}\right) = \frac{1}{2}\log_2\left(\frac{V_{b_i}}{V_{b_i|a_i}}\right) \, , i = 1,\ldots, r \, .
\label{eq25}
\eeq
Similarly, the second term $I\left(Q_{b_i}; E_i\right)$ in (\ref{eq21}) is known as the mutual quantum information between Bob and Eve, which gives a measure of the information that Eve can extract about Bob's measured quadrature mode $Q_{b_i} \in \{X_{b_i}, P_{b_i}\}$. This mutual quantum information $I\left(Q_{b_i};E_i\right)$ is also termed as the Holevo information \cite{ref-7} and is expressed as
\beq
I\left(Q_{b_i};E_i\right) = S \left(E_i \right)
- S \left( E_i \left| Q_{b_i} \right. \! \right) \, ,
\label{eq26}
\eeq
where $S(E_i)$ and $S(E_i|Q_{b_i})$ denote the Von Neumann entropy of Eve's ancillary state and the conditional Von Neumann entropy of Eve's ancillary state conditioned on Bob's quadrature mode, $Q_{b_i}$, respectively. For a given Gaussian state $\tilde{\rho}$ ($=E_i$ or $E_i|Q_{b_i}$) containing two modes, the Von Neumann (or quantum) entropy can be expressed in terms of its symplectic eigenvalues as \cite{weedbrook_gaoussian_van_numen}
\beq
S\left( \tilde{\rho} \right) = \sum_{q=1}^2 h_o \left( \lambda_q \right) \, ,
\label{eq27}
\eeq
where the function $h_o (\cdot)$ outputs the Holevo information given by
\beqarr
h_o (\lambda_q ) \! \! \! \! &=& \! \! \! \! \left(\frac{\lambda_q+1}{2}\right)
\log_2 \left( \frac{\lambda_q+1}{2} \right) \nn \\
&& - \left( \frac{\lambda_q-1}{2} \right)
\log_2 \left( \frac{\lambda_q-1}{2} \right) \, , q=1,2 \, ,
\label{eq28}
\eeqarr
and $\lambda_q$s denote the symplectic eigenvalues of a covariance matrix $\undb{K}_E$, which is of the form given in (\ref{eq18}). In \cite{ neel_spd_channel_estimation_skr, ref-7}, it is studied that a direct computation of the eigenvalues of $\undb{K}_E$ is cumbersome, and thus, the symplectic eigenvalues are obtained by evaluating the magnitudes of the non-zero eigenvalues of the matrix $\mathbf{\Sigma}_E = \jmath\mathbf{\Omega} \undb{K}_E$, with the eigenvalues of both matrices being the same, implying that $\text{eig}\left( \undb{K}_E \right) = \left| \text{eig} \left( \mathbf{\Sigma}_E \right) \right|$. Thus, from (\ref{eq18})-(\ref{eq19b}), and considering the fact that in the $i$-th channel the ancillary state of Eve consists of two modes, $e_{qm_i}$ and $e_{out_i}$, the corresponding matrix $\mathbf{\Sigma}_{E_{j_i}}$ can be expressed as
\beq
\mathbf{\Sigma}_{E_{j_i}} =
\begin{bmatrix}
    V_{e_{out_{j_i}}}\textbf{I}_2 & V_{ee_{j_i}}\textbf{Z}\\
    V_{ee_{j_i}}^* \textbf{Z}^T & V_e \textbf{I}_2
\end{bmatrix}
\, , j \in \left\{ d, g, f \right\} \, , i=1,\ldots, r \, ,
\label{eq29}
\eeq
where $V_{e_{out_{j_i}}}$s and $V_{ee_{j_i}}$s are given in (\ref{eq19a}), and $\textbf{Z}$ is the Pauli-z matrix given as $\text{diag}\left(1,-1  \right)$. This results in the symplectic eigenvalues of the covariance matrix of Eve's ancillary mode are given as \cite{ref-7, Pirandola_2009_lemda_formula}
\beq
\lambda_{j,i_{1,2}} = \sqrt{\frac{1}{2}
\left( \nabla_{j_i} \pm \sqrt{ \nabla^2_{j_i}
-4 \text{det} \left( \mathbf{\Sigma}_{E_{j_i}} \right)} \right)} \, ,
\label{eq30}
\eeq
where
\bsub
\beq
\nabla_{j_i} = 
\begin{cases}
V_{e_{out_{j_i}}}^2 + V_e^2 - 2 \beta_{j_i} \left( {V_e}^2-1 \right)
\, \! \! \! \! &, \ j \in \left\{ d, g \right\} , \\
V_{e_{out_{f_i}}}^2 + V_e^2 - 2 \Re \left\{ \tilde{\beta}_{f_i}^2 \right\} \left( {V_e}^2-1 \right) \,
\! \! \! \! &, \ j = f \, ,
\end{cases}
\label{eq31a}
\eeq
and 
\beq
\text{det} \left(\mathbf{\Sigma}_{E_{j_i}} \right)
\! = \! \begin{cases}
\left(V_{e_{out_{j_i}}} V_e
- \beta_{j_i} \left({V_e}^2-1\right) \right)^2 \, , \,
j \in \left\{ d, g \right\} , \\
\left(V_{e_{out_{f_i}}} V_e
- \left| \tilde{\beta}_{f_i} \right|^2 \left({V_e}^2-1\right) \right)^2 \, , \,
j = f \, .
\end{cases}
\label{eq31b}
\eeq
\esub
Substituting (\ref{eq31a}) and (\ref{eq31b}) in (\ref{eq30}) followed by algebraic simplifications results in the expressions for the symplectic eigenvalues $\lambda_{1,2}$ to be given as in (\ref{eq32}) at the top of this page.

Similarly, the conditional covariance matrix of Eve's state given Bob's quadrature is denoted as $\mathbf{\Sigma}_{E_{j_i}|Q_{b_i}}$ and is defined as \cite{ref-7}
\setcounter{equation}{32}
\beq
\mathbf{\Sigma}_{E_{j_i}|Q_{b_i}} = \mathbf{\Sigma}_{E_{j_i}} - \frac{1}{V_{b_i}}\textbf{W}_{j_i}\textbf{M} \textbf{W}_{j_i}^\dagger \, ,
\label{eq33}
\eeq
where $\mathbf{\Sigma}_{E_{j_i}}$ is given in (\ref{eq29}) and
\beq
\textbf{M} =\begin{bmatrix}
    1&0\\
    0&0
\end{bmatrix} \ , \
\textbf{W}_{j_i} = \begin{bmatrix}
     \langle e_{out_{j_i}} \cdot Q_{b_i} \rangle \textbf{I}_2 \\
     \langle e_{qm_{j_i}}\cdot Q_{b_i}\rangle \textbf{Z}
 \end{bmatrix} .
\label{eq34}
\eeq
From (\ref{eq32}) and (\ref{eq33}), we observe that $\mathbf{\Sigma}_{E_{j_i}|Q_{b_i}}$ can be expressed in the form
\beq
\mathbf{\Sigma}_{E_{j_i}|Q_{b_i}} =
\begin{bmatrix}
   \textbf{A}_{j_i} & \textbf{C}_{j_i} \\
    \textbf{C}^\dagger_{j_i} & \textbf{B}_{j_i}
\end{bmatrix} \, ,
\label{eq35}
\eeq
which is of the same form as (\ref{eq29}) and thus, its symplectic eigenvalues can be computed as
\beq
\lambda_{j,i_{3,4}} =\sqrt{\frac{1}{2}\left( \tilde{\nabla}_{j_i}
\pm \sqrt{\tilde{\nabla}_{j_i}^2
- 4\text{det} \left(\mathbf{\Sigma}_{E_{j_i}|Q_{b_i}} \right)}\right)},
\label{eq36}
\eeq
where
\beq
\tilde{\nabla}_{j_i} = \text{det} \left( \textbf{A}_{j_i} \right)
+ \text{det} \left( \textbf{B}_{j_i} \right)
+ 2 \Re \left\{ \text{det} \left( \textbf{C}_{j_i} \right) \right\} \, .
\label{eq37}
\eeq

\setcounter{equation}{41}
\begin{figure*}[t]
\beqarr
\text{SKR}_i^{\text{RR}} \! \! \! \! &=& \! \! \! \!
\frac{1}{2} \log_2 \left(\frac{\left[ \beta_{d_i} V_a + \left( 1 - \beta_{d_i} \right) V_e \right]
\left[ \beta_{g_i} \beta_{f_i} V_a
+ \left( 1 - \beta_{f_i} \beta_{g_i} + 2 \sqrt{\beta_{f_i} \left(1 - \beta_{f_i} \right) \left(1 - \beta_{g_i} \right)} \cos \phi_i \right) V_e \right]}
{\left[ \beta_{d_i} V_o + \left( 1 - \beta_{d_i} \right) V_e \right]
\left[ \beta_{g_i} \beta_{f_i} V_o
+ \left( 1 - \beta_{f_i} \beta_{g_i} + 2 \sqrt{\beta_{f_i} \left(1 - \beta_{f_i} \right) \left(1 - \beta_{g_i} \right)} \cos \phi_i \right) V_e \right]} \right) \nn \\
&-& \! \! \! \! I \left( Q_{b_{d_i}};E_i \right)
-I\left( Q_{b_{\text{RIS}_i}};E_i \right) \ , \ i=1,\ldots, r
\label{eq42}
\eeqarr
\noindent\rule{\textwidth}{.5pt}
\vspace{-0.8cm}
\end{figure*}
\setcounter{equation}{43}
\begin{figure*}[t]
\begin{eqnarray}
\lambda_{{1,2}_{d_i}} = \sqrt{
\begin{array}{cc}
& \! \! \! \! \! \! \! \! \! \! \! \! \! \! \! \! \!
\! \! \! \! \! \! \! \! \! \! \! \! \! \! \! \! \! \!
\! \! \! \! \! \! \! \! \! \! \! \! \! \! \! \! \! \!
\! \! \! \! \! \!
\frac{1}{2} \left[ \left(1-\beta_{d_i} \right)^2
\left( V_a^2 + V_e^2 \right)
+ 2 \beta_{d_i} \left(1 + \left(1 - \beta_{d_i} \right) V_a V_e \right) \right. \\
& \! \! \! \! \! \! \! \! \left.
\pm \left( 1 - \beta_{d_i} \right) \left(V_a - V_e \right)
\sqrt{\left( \left( 1 - \beta_{d_i} \right) V_a
+ \left( 1 + \beta_{d_i} \right) V_e \right)^2
- 4 \beta_{d_i}\left( V_e^2 - 1 \right)} \right]
\end{array}} \quad , \ i =1,\ldots, r
\label{eq44}
\end{eqnarray}
\noindent\rule{\textwidth}{.5pt}
\vspace{-0.8cm}
\end{figure*}
\begin{figure*}[t]
\setcounter{equation}{46}
\beqarr
\lambda_{{3,4}_{d_i}} = \sqrt{
\frac{\left(1-\beta_{d_i}\right) \left(V_a^2+1\right) V_e + 2 \beta_{d_i} V_a  \pm \left(1-\beta_{d_i}\right) V_e \sqrt{\left(V_a^2+1\right)^2-4 V_e^2}  } 
{2\left( \beta_{d_i} V_a + \left( 1-\beta_{d_i} \right) V_e \right)}} \quad , \ i = 1,\ldots,r
\label{eq47} 
\eeqarr
\noindent\rule{\textwidth}{.5pt}
\vspace{-0.8cm}
\end{figure*}
\begin{figure*}[t]
\beqarr
\text{SKR}_{\text{MIMO}_j}^{\text{RR}} \! \! \! \! &=& \! \! \! \! 
\frac{1}{2} \sum_{i=1}^r \log_2 \left(\frac{\left[ \beta_{d_i} V_a + \left( 1 - \beta_{d_i} \right) V_e \right]
\left[ \beta_{g_i} \beta_{f_i} V_a
+ \left( 1 - \beta_{f_i} \beta_{g_i} + 2 \sqrt{\beta_{f_i} \left(1 - \beta_{f_i} \right) \left(1 - \beta_{g_i} \right)} \cos \phi_i \right) V_e \right]}
{\left[ \beta_{d_i} V_o + \left( 1 - \beta_{d_i} \right) V_e \right]
\left[ \beta_{g_i} \beta_{f_i} V_o
+ \left( 1 - \beta_{f_i} \beta_{g_i} + 2 \sqrt{\beta_{f_i} \left(1 - \beta_{f_i} \right) \left(1 - \beta_{g_i} \right)} \cos \phi_i \right) V_e \right]} \right)  \nn \\
&& \! \! \! \! - \sum_{i=1}^r
\left[ h_o \left( \lambda_{1_{j_i}} \right)
+ h_o \left( \lambda_{2_{j_i}} \right)
- h_o \left( \lambda_{3_{j_i}} \right)
- h_o \left( \lambda_{4_{j_i}} \right) \right]
\ , \ j \in \left\{ d, g, f \right\}
\label{eq48}
\eeqarr
\noindent\rule{\textwidth}{.5pt}
\vspace{-0.8cm}
\end{figure*}

From (\ref{eq20}), the SKR for $i$-th parallel channel of the direct path between the transceiver pair can be expressed as
\setcounter{equation}{37}
\beq
\left( \Delta I \right)_{\text{LoS}_i} = \frac{1}{2} \log_2 \left( \frac{V_{b_{d_i}}}{V_{b_{d_i}|a_i}} \right) -I\left( Q_{b_{d_i}};E_i \right),
\label{eq38}
\eeq
and similarly, the SKR for the equivalent $i$-th parallel channel via the RIS can be depicted by
\beq
\left(\Delta I\right)_{\text{RIS}_i}=\frac{1}{2}\log_2\left(\frac{V_{b_{\text{RIS}_i}}}{V_{b_{\text{RIS}_i}|a_i}}\right) -I\left(Q_{b_{\text{RIS}_i}};E_i\right),
\label{eq39}
\eeq
where $V_{b_{d_i}}$s and $V_{b_{\text{RIS}_i}}$s are given in (\ref{eq16a}), and we have
\beqarr
V_{b_{d_i}|a_i} \! \! \! \! &=& \! \! \! \!
\beta_{d_i} V_o + \left(1-\beta_{d_i}\right) V_e \, , \nn\\
V_{b_{{\text{RIS}}_i}|a_i} \! \! \! \! &=& \! \! \! \!
\left( \alpha_i \alpha_i ^* \right) V_o 
+ \left( \gamma_i \gamma_i^* \right) V_e \, , \, i = 1,\ldots, r \, ,
\label{eq40}
\eeqarr
with $\alpha_i$s and $\gamma_i$s are given in (\ref{eq16b}). Thus using (\ref{eq38}) and (\ref{eq39}), the final expression for SKR for the $i$-th channel is obtained as
\beqarr
&& \! \! \! \! \! \! \! \! \! \! \! \! \! \! \! \! \! \! \! \!
\! \! \! \! \! \! \! \!
\text{SKR}_i^{\text{RR}}
=\frac{1}{2} \log_2 \left( \frac{V_{b_{d_i}}V_{b_{\text{RIS}_i}}}{V_{b_{d_i}|a_i}V_{b_{\text{RIS}_i}|a_i}} \right) \nn \\
&& \! \! \! \! \! \! \! \!
- I \left( Q_{b_{d_i}};E_i \right)-I\left( Q_{b_{\text{RIS}_i}};E_i \right) \, , \, i=1,\ldots, r \, .
\label{eq41}
\eeqarr
Substituting (\ref{eq16a}), (\ref{eq16b}), and (\ref{eq40}) in (\ref{eq41}) followed by algebraic simplifications results in the expression for the SKR in the $i$-th channel to be given as in (\ref{eq42}) at the top of this page.
Moreover, the effective secret key rate of the RIS-assisted MIMO CV-QKD Gaussian-modulated system with the presence of a direct link can be expressed as
\setcounter{equation}{42}
\beq
\text{SKR}_{\text{MIMO}}^{\text{RR}}
=\sum_{i=1}^r \text{SKR}_{i}^{\text{RR}} \, .
\label{eq43}
\eeq
It is to be noted that for the secure transmission of a key over the channel, $\text{SKR}_{\text{MIMO}}^{\text{RR}} > 0$. Furthermore, the formulation presented has been carried out for a general case of the system under study. Recall that we have considered a limited quantum memory of Eve, implying that Eve has the limited ability to store the ancilla modes from one of the wireless channels in the system. Thus, in the following subsections, we consider three cases of the measurement via one of the wireless channels and compute the expression of the SKR. 
\subsection{Measurement of Ancillary Mode $\{e_{out_{d}},e_{qm}\}$}
We first consider the case of Eve storing the ancillary mode for the measurement from the direct path between the transceiver pair. For this case, the value of $I\left(Q_{b_{\text{RIS}_i}}; E_i\right)$ in (\ref{eq42}) is equal to zero due to the lack of ancillary mode measurement through the RIS path. Thus, to compute the expression of SKR, we need to obtain the expression for $I\left(Q_{b_{d_i}}; E_i\right)$ for which the expressions for the eigenvalues $\lambda_{1,2}$ and $\lambda_{3,4}$ need to be computed. Substituting the expression of $V_{e_{out_{d_i}}}$ from (\ref{eq17}) in (\ref{eq32}) followed by algebraic simplifications results in the expression for $\lambda_{1,2}$ to be obtained as in (\ref{eq44}) at the top of this page.
Similarly, to compute $\lambda_{3,4}$, the matrix elements of the conditional covariance matrix in (\ref{eq35}) is given by
\setcounter{equation}{48}
\begin{figure*}
\begin{eqnarray}
\lambda_{{1,2}_{g_i}} = \sqrt{
\begin{array}{cc}
& \! \! \! \! \! \! \! \! \! \! \! \! \! \! \! \! \!
\! \! \! \! \! \! \! \! \! \! \! \! \! \! \! \! \! \!
\! \! \! \! \! \! \! \! \! \! \! \! \! \! \! \! \! \!
\! \! \! \! \! \!
\frac{1}{2} \left[ \left(1-\beta_{g_i} \right)^2
\left( V_a^2 + V_e^2 \right)
+ 2 \beta_{g_i} \left(1 + \left(1 - \beta_{g_i} \right) V_a V_e \right) \right. \\
& \! \! \! \! \! \! \! \! \left.
\pm \left( 1 - \beta_{g_i} \right) \left(V_a - V_e \right)
\sqrt{\left( \left( 1 - \beta_{g_i} \right) V_a
+ \left( 1 + \beta_{g_i} \right) V_e \right)^2
- 4 \beta_{g_i}\left( V_e^2 - 1 \right)} \right]
\end{array}} \quad , \quad i =1,\ldots, r
\label{eq49}
\end{eqnarray}
\noindent\rule{\textwidth}{.5pt}
\end{figure*}
\begin{figure*}[t]
\beqarr
\textbf{A}_{g_i} \! \! \! \! &=& \! \! \! \!
\text{diag} \left( \frac{\left( 1-\beta_{g_i} + \beta_{g_i} \beta_{f_i} + 2\sqrt{\beta_{\!f_i} \!\left(\!1\! -\! \beta_{\!f_{i}}\! \right) \!\left(\!1\! -\! \beta_{g_i}\! \right)}\cos \phi_i \right)V_a V_e}
{\beta_{g_i} \beta_{f_i} V_a + \left(1- \beta_{g_i} \beta_{f_i} +2\sqrt{\beta_{\!f_i} \!\left(\!1\! -\! \beta_{\!f_{i}}\! \right) \!\left(\!1\! -\! \beta_{g_i} \right)} \cos \phi_i \right)V_e} , 
\left( 1-\beta_{g_i}\right) V_a + \beta_{g_i}V_e \right), \nn \\
\textbf{B}_{g_i} \! \! \! \!  &=& \! \! \! \!
\text{diag} \left( \frac{ \left( 1- \beta_{g_i} \beta_{f_i} +2\sqrt{\beta_{\!f_i} \!\left(\!1\! -\! \beta_{\!f_{i}}\! \right) \!\left(\!1\! -\! \beta_{g_i} \right)} \cos \phi_i +\beta_{g_i} \beta_{f_i} V_a V_e \right)}
{\beta_{g_i} \beta_{f_i} V_a + \left(1- \beta_{g_i} \beta_{f_i} +2\sqrt{\beta_{\!f_i} \!\left(\!1\! -\! \beta_{\!f_{i}}\! \right) \!\left(\!1\! -\! \beta_{g_i} \right)} \cos \phi_i \right)V_e}, V_e \right) , \nn \\
\textbf{C}_{g_i} \! \! \! \! &=& \! \! \! \!
\text{diag} \left( \frac{\left( \beta_{f_i} + \sqrt{ \beta_{f_i} \left(  1-\beta_{g_i} \right) \left( 1-\beta_{f_i}\right) }e^{-\jmath \phi} \right) \sqrt{\beta_{g_i}\left(V_e^2-1\right) }V_a}
{\beta_{g_i} \beta_{f_i} V_a + \left(1- \beta_{g_i} \beta_{f_i} +2\sqrt{\beta_{\!f_i} \!\left(\!1\! -\! \beta_{\!f_{i}}\! \right) \!\left(\!1\! -\! \beta_{g_i} \right)} \cos \phi_i \right)V_e},
- \sqrt{\beta_{g_i}\left({V_e}^2-1\right)} \right)
\label{eq50}
\eeqarr
\noindent\rule{\textwidth}{.5pt}
\end{figure*}
\begin{figure*}
\beqarr
&& \! \! \! \! \! \! \! \! \! \! \! \! \! \! \! \! \! \! \! \! \!
\tilde{\nabla}_{g_i} =
\frac{ \left[ \begin{array}{ll}
& \! \! \! \! \! \! \! \!
\left( 1-\beta_{g_i} \right) \left(1-\beta_{g_i} + \beta_{g_i} \beta_{f_i} +2\sqrt{\beta_{f_i} \left(1 -\beta_{f_{i}} \right) \left(1 -\beta_{g_i} \right)} \cos \phi_i\right) V_a^2 V_e
+ \left(1-\beta_{g_i}- \beta_{f_i}+\beta_{g_i}\beta_{f_i}\right) 
   \beta_{g_i}V_a V_e^2 \\
& \! \! \! \! \! \! \! \!
+ 2 \left(\beta_{f_{i}}+ \sqrt{\beta_{f_i} \left(1 -\beta_{f_{i}} \right) 
\left(1 - \beta_{g_i} \right)} \cos \phi_i\right) \beta_{g_i} V_a
+ \left(1-\beta_{g_i} \beta_{f_i} +2\sqrt{\beta_{f_i} \left(1 - 
  \beta_{f_{i}} \right) \left(1 -\beta_{g_i} \right)} \cos \phi_i\right) V_e \end{array} \right] }
{\beta_{g_i} \beta_{f_i} V_a + \left(1- \beta_{g_i} \beta_{f_i} +2\sqrt{\beta_{f_i} \left(1 -\beta_{f_{i}} \right) \left(1 - \beta_{g_i} \right)} \cos \phi_i \right)V_e} , \nn \\
&& \! \! \! \! \! \! \! \! \! \! \! \! \! \! \! \! \! \! \! \! \!
\text{det} \left(\mathbf{\Sigma}_{E_{g_i}|Q_{b_{\text{RIS}_i}}} \right) \nn \\
&& \! \! \! \! \! \! \! \! \! \! 
= \frac{ \left[ \begin{array}{ll}
& \! \! \! \! \! \! \! \! \! \! 
\left(\left(1-\beta_{g_i} \right) V_a V_e
+ \beta_{g_i}\right) \left( \left( \left(1-\beta_{g_i} \beta_{f_i} \right) 
\left( 1-\beta_{g_i}+ \beta_{g_i} \beta_{f_i} \right)
+ 2\left( 2\beta_{f_i} \left(1 - \beta_{f_i} \right) \left(1 - \beta_{g_i} \right) 
+ \left(2-\beta_{g_i}\right) \right. \right. \right. \\
& \! \! \! \! \! \! \! \! \! \! 
\left. \left. \times \sqrt{\beta_{f_i} \left(1 - \beta_{f_i} \right) \left(1 - \beta_{g_i} \right)} \right) \cos \phi_i \right) V_a V_e
- \left( \beta_{g_i} \beta_{f_i} V_a V_e\right)^2
+\left( \beta_{f_i} \left( 1+\beta_{g_i}+\beta_{g_i} \beta_{f_i}\right) \right. \\
& \! \! \! \! \! \! \! \! \! \! 
\left. \left. +2\beta_{g_i} \sqrt{\beta_{f_i} \left(1 - \beta_{f_i} \right) \left(1 - \beta_{g_i} \right)} \cos \phi_i \right) \beta_{g_i} V_a^2 \right)
\end{array} \right]}
{\left(\beta_{g_i} \beta_{f_i} V_a + \left(1- \beta_{g_i} \beta_{f_i} +2\sqrt{\beta_{f_i} \left(1 - \beta_{f_i} \right) \left(1 - \beta_{g_i} \right)} \cos \phi_i \right)V_e \right)^2}
\label{eq51}
\eeqarr
\noindent\rule{\textwidth}{.5pt}
\vspace{-0.8cm}
\end{figure*}
\begin{figure*}
\begin{small}
\begin{eqnarray}
\lambda_{{1,2}_{f_i}} =
\sqrt{ \frac{1}{2} \left[
\begin{array}{ccc}
& \! \! \! \! \! \! \! \! \! \! \! \! \! \! \! \!
\left( \left(1-\beta_{f_i}\right) \beta_{g_i} V_a 
+ \left(1- \beta_{g_i} + \beta_{g_i} \beta_{f_i}  - 2 \sqrt{\beta_{f_i} \left(1 - \beta_{f_i} \right) \left(1 - \beta_{g_i}\right)} \cos{\phi_i} \right) V_e \right)^2 + V_e^2\\
& \!\!\!\!\! \! \! \! \! \! \! \! \!\! \! \! \! \! \!
\! \! \! \!\! \! \! \!  \! \! \! \! \! \! \! \! \!
- 2 \left( \beta_{f_i}+ \left(1-\beta_{g_i} \right)\left(1-\beta_{f_i} \right) \cos \left(2\phi_i\right) -2 \sqrt{\beta_{f_i} \left(1 - \beta_{f_i} \right) \left(1 - \beta_{g_i} \right)} \cos \phi_i \right) \\
& \! \! \! \! \! \! \! \!
\pm \sqrt{ \begin{array}{ll}
& \! \! \! \! \! \! \! \!
\left( \left( \left(1-\beta_{f_i}\right) \beta_{g_i} V_a
+ \left(2- \beta_{g_i} + \beta_{g_i} \beta_{f_i} - 2 \sqrt{\beta_{f_i} \left(1 - \beta_{f_i} \right) \left(1 - \beta_{g_i}\right)} \cos{\phi_i} \right) V_e \right)^2 \right. \\
& \! \! \! \! \! \! \! \!
-4 \left( \beta_{f_i}+ \left(1-\beta_{g_i} \right) \left(1-\beta_{f_i} \right) \cos^2 \phi_i -2 \sqrt{\beta_{f_i} \left(1 - \beta_{f_i} \right) \left(1 - \beta_{g_i} \right)} \cos \phi_i \right) \left(V_e^2-1 \right) \\
& \! \! \! \! \! \! \! \!  \times
\left( \left( \left(1-\beta_{f_i}\right) \beta_{g_i} V_a 
+ \left(\beta_{g_i} \beta_{f_i} -\beta_{g_i} - 2 \sqrt{\beta_{f_i} \left(1 - \beta_{f_i} \right) \left(1 - \beta_{g_i}\right)} \cos{\phi_i} \right) V_e \right)^2 \right. \\
& \! \! \! \! \! \! \! \!
-4 \left( \left(1-\beta_{g_i} \right) \left(1-\beta_{f_i} \right) \sin^2 \phi_i \left(V_e^2-1 \right) \right)
\end{array}}
\end{array} \right]} \, , i =1,\ldots, r
\label{eq52}
\vspace{-0.8cm}
\end{eqnarray}
\end{small}
\noindent\rule{\textwidth}{.5pt}
\vspace{-0.9cm}
\end{figure*}
\setcounter{equation}{44}
\beqarr
\textbf{A}_{d_i} \! \! \! \! &=& \! \! \! \!
\text{diag} \left( \frac{V_a V_e}
{\beta_{d_i} V_a + \left(1 - \beta_{d_i} \right) V_e} ,
\left(1-\beta_{d_i}\right) V_a + \beta_{d_i} V_e \right), \nn \\
\textbf{B}_{d_i} \! \! \! \! &=& \! \! \! \!
\text{diag} \left( \frac{ \left(1-\beta_{d_i}
+ \beta_{d_i} V_a V_e \right)}
{\beta_{d_i} V_a + \left(1-\beta_{d_i} \right) V_e}, V_e \right) , \nn \\
\textbf{C}_{d_i} \! \! \! \! &=& \! \! \! \!
\text{diag} \left( \frac{V_aV_{ed_i}}
{\beta_{d_i} V_a + \left(1 - \beta_{d_i} \right) V_e},
- \sqrt{\beta_{d_i}\left({V_e}^2-1\right)} \right) . \nn \\
\label{eq45}
\eeqarr
Substituting (\ref{eq45}) in (\ref{eq37}) followed by algebraic simplifications results in
\bsub
\beq
\tilde{\nabla}_{d_i} = \frac{ \left( 1-\beta_{d_i} \right) \left( V_a^2+1 \right) V_e + 2 \beta_{d_i}V_a} 
{\beta_{d_i} V_a + \left( 1-\beta_{d_i} \right) V_e} \, ,
\label{eq46a}
\eeq
and 
\beqarr
&& \! \! \! \! \! \! \! \! \! \! \! \! \! \! \! \! \! \! \! \!
\! \! \! \!
\text{det} \left(\mathbf{\Sigma}_{E_{d_i}|Q_{b_{d_i}}} \right) \nn \\
&& \! \! \! \! \! \! \! \! \! \! \! \! \! \! \! \! \! \! \! \!
= \frac{\left(\left(1-\beta_{d_i} \right) V_a V_e
+ \beta_{d_i}\right) \left( \left(1-\beta_{d_i} \right) V_a V_e
+ \beta_{d_i} V_a^2\right)}
{\left( \beta_{d_i} V_a + \left( 1-\beta_{d_i} \right) V_e \right)^2}
\, .
\label{eq46b}
\eeqarr
\esub
Using (\ref{eq46a}) and (\ref{eq46b}) and simplifying further, we obtain the expressions for $\lambda_{3,4}$ as given in (\ref{eq47}) at the top of the previous page. Moreover, the substitution of (\ref{eq42}) in (\ref{eq41}) results in the expression of the SKR for the case where the measurement of ancillary mode, $\{e_{out_{d}},e_{qm}\}$ is performed, to be expressed as given in (\ref{eq48}) for $j=d$ at the top of the previous page, where the eigenvalues are given in (\ref{eq44}) and (\ref{eq45}) and the function $h_o(\cdot)$ is defined in (\ref{eq28}).
\setcounter{equation}{52}
\begin{figure*}[t]
\begin{small}
\beqarr
\textbf{A}_{f_i} \! \! \! \! \! &=& \! \! \! \! \!
\text{diag} \left( \frac{\left(\beta_{g_i} V_a V_e + 4\left( 
\beta_{g_i} -\beta_{f_i} \right)\cos^2 \phi_i V_e^2 \right)}
{\beta_{g_i} \beta_{f_i} V_a + \left(1- \beta_{g_i} \beta_{f_i} +2\sqrt{\beta_{\!f_i} \!\left(\!1\! -\! \beta_{\!f_{i}}\! \right) \!\left(\!1\! -\! \beta_{g_i} \right)} \cos \phi_i \right)V_e} , \right. \nn \\
&& \qquad \qquad \left. \beta_{g_i} \left( 1-\beta_{f_i}\right) V_a
+ \left(1-\beta_{g_i} + \beta_{g_i} \beta_{f_i}- 2\sqrt{\beta_{f_i} \left(1 - \beta_{f_{i}} \right) \left(1 - \beta_{g_i} \right)} \cos \phi_i \right)V_e \right), \nn \\
\textbf{B}_{f_i} \! \! \! \! \! &=& \! \! \! \! \!
\text{diag} \left( \frac{ \left( \beta_{g_i} \beta_{f_i}  V_a V_e + 1-\beta_{g_i} \beta_{f_i}+2\sqrt{\beta_{f_i} \left(1 - \beta_{f_{i}} \right) \left(1 - \beta_{g_i} \right)} \cos \phi_i \right)}
{\beta_{g_i} \beta_{f_i} V_a + \left(1- \beta_{g_i} \beta_{f_i} +2\sqrt{\beta_{f_i} \left(1 - \beta_{f_{i}} \right) \left(1 - \beta_{g_i} \right)} \cos \phi_i \right)}, V_e \right) , \nn \\
\textbf{C}_{f_i} \! \! \! \! \! &=& \! \! \! \! \!
\text{diag} \left( \frac{\left( \left(1+\sqrt{\beta_{f_i} \left(1 - \beta_{f_{i}} \right) \left(1 - \beta_{g_i}\right)} \left( 1- e^{\jmath \phi_i}\right) \right)\beta_{g_i} V_a 
+\left(1-2\cos \phi_i e^{\jmath \phi_i} + e^{\jmath 2\phi_i} \right) \left(1-\beta_{g_i}\right) V_e \right) 
\sqrt{\beta_{f_i}\left(V_e^2-1\right) }}
{\beta_{g_i} \beta_{f_i} V_a + \left(1- \beta_{g_i} \beta_{f_i} +2\sqrt{\beta_{f_i} \left(1 - \beta_{f_{i}} \right) \left(1 - \beta_{g_i} \right)} \cos \phi_i \right)V_e} , \right. \nn \\
&& \qquad \qquad \left. -\left(\sqrt{\beta_{f_i}} - \sqrt{\left(1 - \beta_{g_i}\right) \left(1 - \beta_{f_i} \right)} e^{\jmath \phi_i}  \right) \sqrt{\left({V_e}^2-1\right)} \right)
\label{eq53}
\eeqarr
\end{small}
\noindent\rule{\textwidth}{.5pt}
\vspace{-0.8cm}
\end{figure*}
\begin{figure*}
\begin{small}
\beqarr
&& \! \! \! \! \! \! \! \! \!
\tilde{\nabla}_{f_i} = \nn \\
&& \! \! \! \! \! \! \! \! \!
\frac{ \left[ \begin{array}{ll}
& \! \! \! \! \! \! \! \!
\beta_{g_i}^2 \left( 1 \! - \! \beta_{f_i} \right) \! V_a^2 V_e
\! + \! \left( 4\left( 1 - \beta_{f_i} \right)
\left( \beta_{g_i} - \beta_{f_i} \right) \cos^2 \phi_i
\! + \! \left( 1 \! - \! \beta_{g_i} \! + \! \beta_{f_i} \! + \! \beta_{g_i} \beta_{f_i}
\! - \! 2 \sqrt{\beta_{f_i} \left(1 - \beta_{f_{i}} \right)
\left(1 - \beta_{g_i} \right)} \cos \phi_i \right) \right)
\beta_{g_i} V_a V_e^2 \\
& \! \! \! \! \! \! \! \!
+ \left( 1 - \beta_{g_i} \beta_{f_i}
+ 2\sqrt{\beta_{f_i} \left(1 - \beta_{f_{i}} \right)
\left(1 - \beta_{g_i} \right)} \cos \phi_i \right) V_e 
+ 4 \left( \beta_{g_i} - \beta_{f_i} \right)
\left( 1 - \beta_{g_i} + \beta_{g_i} \beta_{f_i}
- 2 \sqrt{\beta_{f_i} \left(1 - \beta_{f_{i}} \right)
\left(1 - \beta_{g_i} \right)} \cos \phi_i \right) \\
& \! \! \! \! \! \! \!
\times \cos^2 \phi_i V_e^3
-2 \left( \beta_{g_i} \left(1+\sqrt{\beta_{f_i} \left(1 -\beta_{f_{i}} \right)
\left(1 - \beta_{g_i} \right)} \left(1- \cos \phi_i \right) \right) 
- \sqrt{\beta_{f_i} \left(1-\beta_{f_{i}}\right) \left(1-\beta_{g_i} \right)} 
\right. \\
& \! \! \! \! \! \! \left.
\times \left( \cos \phi_i -\sqrt{\beta_{f_i} \left(1 -\beta_{f_{i}} \right)
\left(1 - \beta_{g_i} \right)} \left(\cos \left(2\phi_i \right)
-\cos \phi_i \right) \right) \right)
\beta_{g_i} V_a \left(V_e^2-1\right)
\end{array} \right] } 
{\beta_{g_i} \beta_{f_i} V_a + \left(1- \beta_{g_i} \beta_{f_i} +2\sqrt{\beta_{f_i} \left(1 -\beta_{f_{i}} \right) \left(1 - \beta_{g_i} \right)} \cos \phi_i \right)V_e} , \nn \\
&& \! \! \! \! \! \!
\text{det} \left(\mathbf{\Sigma}_{E_{f_i}|Q_{b_{\text{RIS}_i}}} \right)
\nn \\
&& \qquad
= \frac{ \left[ \begin{array}{ll}
& \! \! \! \! \! \! \! \! \! \! 
\left(\beta_{g_i} \left(1-\beta_{f_i} \right) V_a V_e
+ 1-\beta_{g_i}+\beta_{g_i}\beta_{f_i} - 2\sqrt{\beta_{f_i} \left(1 -\beta_{f_{i}} \right) \left(1 - \beta_{g_i} \right)} \cos \phi_i \right) \\
& \!\!\!\!\!\!\!
\times \left( \left(\beta_{g_i} \beta_{f_i} V_a V_e + 1-\beta_{g_i}\beta_{g_i}+2 \sqrt{\beta_{f_i} \left(1 -\beta_{f_{i}} \right) \left(1 - \beta_{g_i} \right)} \cos \phi_i \right) 
\left(\beta_{g_i} V_a V_e +4\left(\beta_{g_i}-\beta_{f_i}\right) \cos^2 \phi_i V_e^2\right) \right. \\
&\!\!\!\!\!\! \! \! 
- \left( \left( 1+2\sqrt{\beta_{f_i} \left(1 -\beta_{f_{i}} \right) \left(1 - \beta_{g_i} \right)} \left(1-\cos \phi_i \right) \left(1+\sqrt{\beta_{f_i} \left(1 -\beta_{f_{i}} \right) \left(1 - \beta_{g_i} \right)}  \right) \right) \left(\beta_{g_i} V_a \right)^2 \right.  \\
& \! \! \! \! \! \! \! \! \left. \left.
+ 4 \cos \phi_i \left( \cos \phi_i -1 \right) \left(1-\beta_{g_i} \right)^2 V_e^2 \right) \beta_{g_i} \left(V_e^2 -1\right) \right)
\end{array} \right]}
{\left(\beta_{g_i} \beta_{f_i} V_a + \left(1- \beta_{g_i} \beta_{f_i} +2\sqrt{\beta_{f_i} \left(1 - \beta_{f_i} \right) \left(1 - \beta_{g_i} \right)} \cos \phi_i \right)V_e \right)^2} 
\label{eq54}
\eeqarr
\end{small}
\noindent\rule{\textwidth}{.5pt}
\vspace{-0.8cm}
\end{figure*}
\subsection{Measurement of Ancillary Mode, $\{e_{out_{g}},e_{qm}\}$}
We consider the case of Eve storing the measurement arising from the data transmitted between the transmitter and the RIS, implying that the term $I\left(Q_{b_{d_i}}; E_i\right)$ in (\ref{eq42}) becomes zero. For this case, we denote the SKR by $\text{SKR}_{\text{MIMO}_g}^{\text{RR}}$, whose computation requires the expression for $I\left(Q_{b_{\text{RIS}_i}}; E_i\right)$, which further implies that we need to find the corresponding expressions for $\lambda_{1,2}$ and $\lambda_{3,4}$. Substituting the expression of $V_{e_{out_{g_i}}}$ from (\ref{eq17}) in (\ref{eq32}) followed by algebraic simplifications results in the expression for $\lambda_{1,2}$ to be obtained as in (\ref{eq49}) given at the top of this page. Similarly, to compute $\lambda_{3,4}$, the matrix elements of the conditional covariance matrix in (\ref{eq35}) are expressed in (\ref{eq50}) at the top of this page.

Substituting (\ref{eq50}) in (\ref{eq37}) followed by algebraic simplifications results in the expression of $\tilde{\nabla}_{g_i}$ and $\text{det} \left(\mathbf{\Sigma}_{E_{g_i}|Q_{b_{\text{RIS}_i}}} \right)$ to be obtained as in (\ref{eq51}) at the top of the previous page, using which the expressions for $\lambda_{3,4}$ can be obtained. It is to be noted that these eigenvalues aren't explicitly expressed here owing to their mathematical complexity and thus, using the appropriate substitutions, we obtain the expression for the SKR for this case from (\ref{eq48}) for $j=g$, with $h_o(\cdot)$ defined in (\ref{eq28}).
\subsection{Measurement of Ancillary Mode, $\{e_{out_{f}},e_{qm}\}$}
Finally, we consider the case of Eve storing the measurement arising from the data transmitted between the RIS and the receiver, implying that the term $I\left(Q_{b_{d_i}}; E_i\right)$ in (\ref{eq42}) becomes zero. For this case, we denote the SKR by $\text{SKR}_{\text{MIMO}_f}^{\text{RR}}$, whose computation requires the expression for $I\left(Q_{b_{\text{RIS}_i}}; E_i\right)$, which further implies that we need to find the corresponding expressions for $\lambda_{1,2}$ and $\lambda_{3,4}$. Substituting the expression of $V_{e_{out_{f_i}}}$ from (\ref{eq17}) in (\ref{eq32}) followed by algebraic simplifications results in the expression for $\lambda_{1,2}$ to be obtained as in (\ref{eq52}) given at the top of the previous page.
Similarly, to compute $\lambda_{3,4}$, the matrix elements of the conditional covariance matrix in (\ref{eq35}) are expressed in (\ref{eq53}) at the top of this page.

Substituting (\ref{eq53}) in (\ref{eq37}) followed by algebraic simplifications results in the expression of $\tilde{\nabla}_{f_i}$ and $\text{det} \left(\mathbf{\Sigma}_{E_{f_i}|Q_{b_{\text{RIS}_i}}} \right)$ to be obtained as in (\ref{eq54}) at the top of this page, using which the expressions for $\lambda_{3,4}$ can be obtained. It is to be noted that these eigenvalues aren't explicitly expressed here owing to their mathematical complexity and thus, using the appropriate substitutions, we obtain the expression for the SKR for this case from (\ref{eq48}) for $j=f$, with $h_o(\cdot)$ defined in (\ref{eq28}).
\section{Numerical Results}
\begin{figure*}
     \centering
     \begin{subfigure}[b]{0.334\textwidth}
         \centering
        \includegraphics[width=\textwidth]{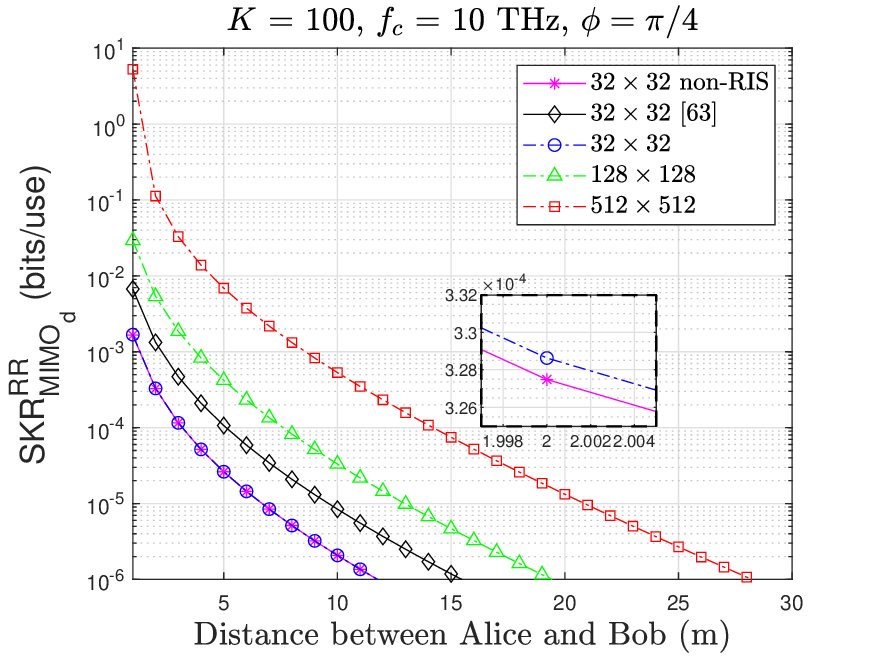}
         \caption{Ancillary mode $\left\{ e_{out_d} , e_{qm} \right\}$}
         \label{fig;3a}
     \end{subfigure}
     \hspace{-0.3cm}
     \begin{subfigure}[b]{0.334\textwidth}
         \centering
        \includegraphics[width=\textwidth]{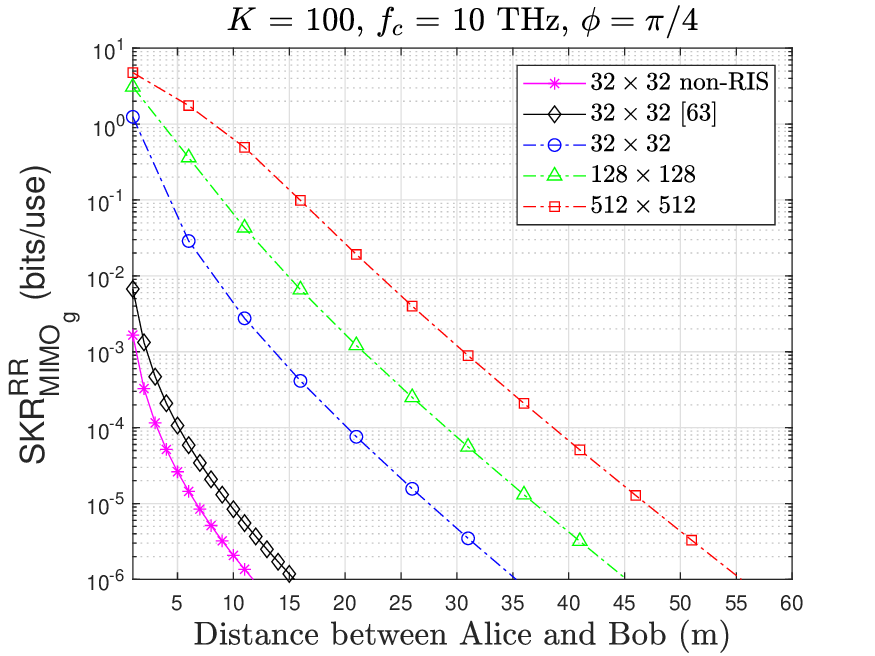}
         \caption{Ancillary mode $\left\{ e_{out_g} , e_{qm} \right\}$}
         \label{fig;3b}
     \end{subfigure}
     \hspace{-0.3cm}
     \begin{subfigure}[b]{0.334\textwidth}
         \centering
        \includegraphics[width=\textwidth]{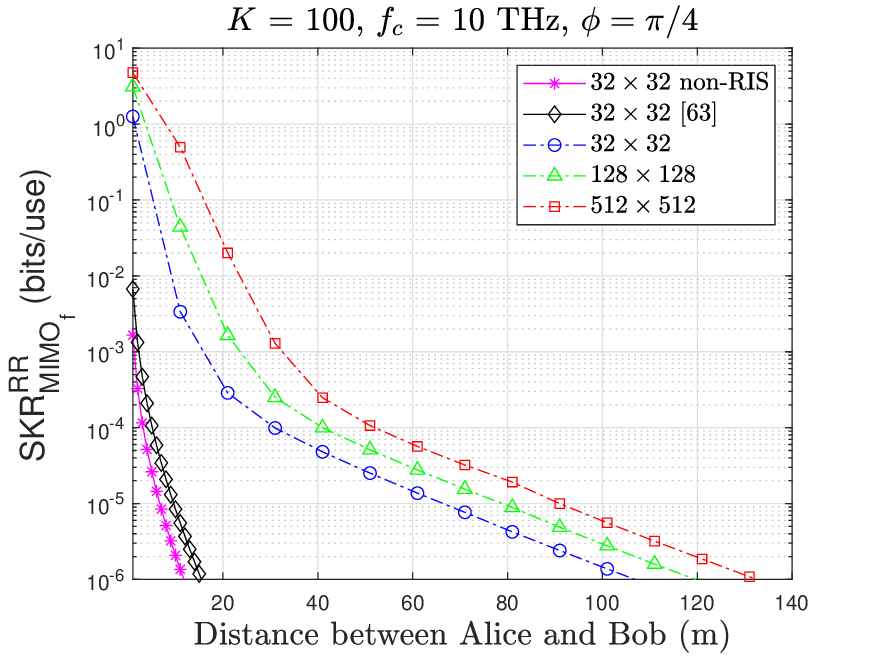}
         \caption{Ancillary mode $\left\{ e_{out_f} , e_{qm} \right\}$}
         \label{fig;3c}
     \end{subfigure}
        \caption{SKR versus distance between Alice and Bob for $N_{R_X}=N_{T_X}=32,128,512$, $K$ = 100, $f_c$ = 10 THz, $\phi=\pi/4$, and $d_a=0.5\lambda_c$ for Eve storing the ancilla modes (a) $\{e_{out_d},e_{qm}\}$, (b) $\{e_{out_g},e_{qm}\}$, and (c) $\{e_{out_f},e_{qm}\}$.}
        \label{f3}
        \vspace{-0.3cm}
\end{figure*}

\begin{figure*}
     \centering
     \begin{subfigure}[b]{0.334\textwidth}
         \centering
        \includegraphics[width=\textwidth]{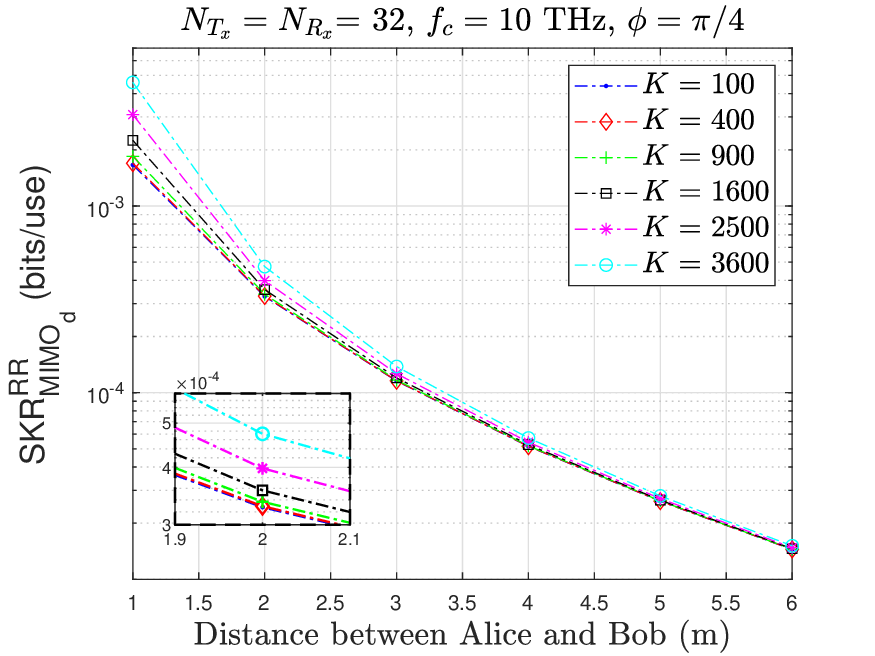}
         \caption{Ancillary mode $\left\{ e_{out_d} , e_{qm} \right\}$}
         \label{fig;4a}
     \end{subfigure}
     \hspace{-0.3cm}
     \begin{subfigure}[b]{0.334\textwidth}
         \centering
        \includegraphics[width=\textwidth]{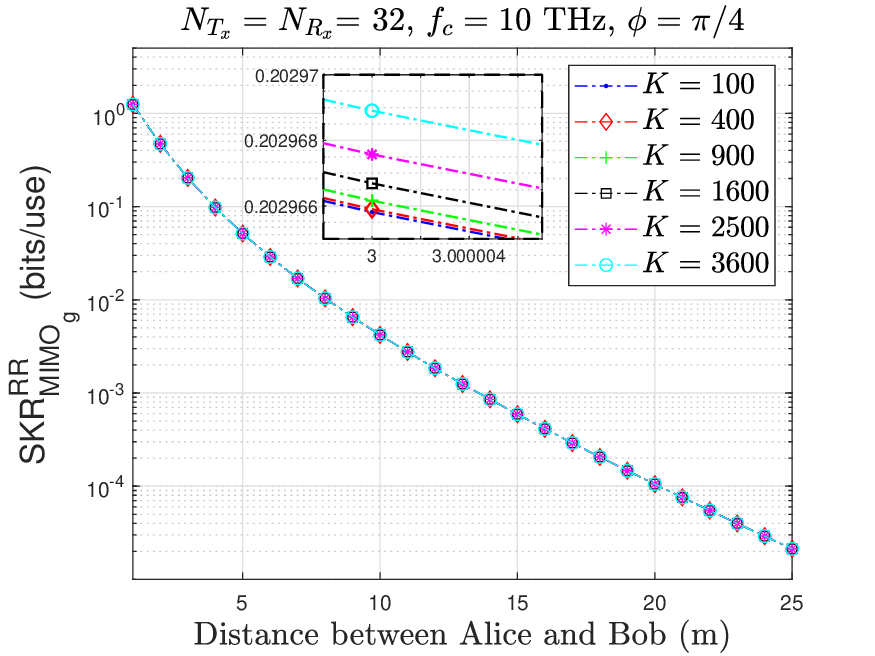}
         \caption{Ancillary mode $\left\{ e_{out_g} , e_{qm} \right\}$}
         \label{fig;4b}
     \end{subfigure}
     \hspace{-0.3cm}
     \begin{subfigure}[b]{0.334\textwidth}
         \centering
        \includegraphics[width=\textwidth]{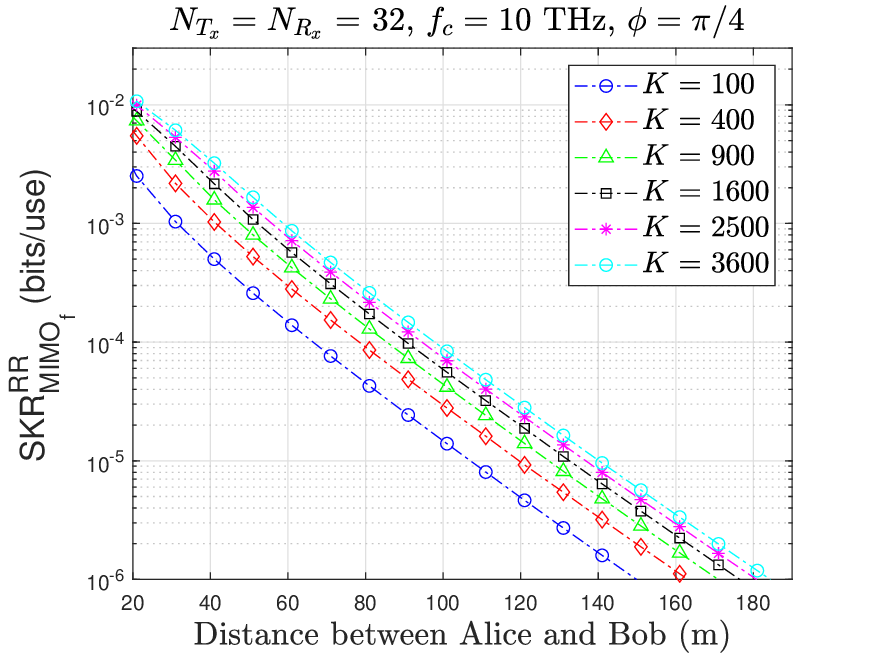}
         \caption{Ancillary mode $\left\{ e_{out_f} , e_{qm} \right\}$}
         \label{fig;4c}
     \end{subfigure}
        \caption{SKR versus distance between Alice and Bob for $N_{R_X}=N_{T_X}=32$, $K = 100, 400, 900, 1600, 2500, 3600$, $f_c$ = 10 THz, $\phi=\pi/4$, and $d_a=0.5\lambda_c$ for Eve storing the ancilla modes (a) $\{e_{out_d},e_{qm}\}$, (b) $\{e_{out_g},e_{qm}\}$, and (c) $\{e_{out_f},e_{qm}\}$.}
        \label{f4}
        \vspace{-0.3cm}
\end{figure*}
This section presents the numerical results corroborating the analytical framework described and derived in the paper. For the simulation studies, we consider the standard system parameters as $\rho$ = 1000 dB/Km, $T_e$ = 300 K (denoting the room temperature), antenna gain $G_a$ = 30 dBi, the variance of Alice's initial modulated signal $V_s = 1000$, the variance of the vacuum state $V_o=2\Bar{n}+1$ with $\Bar{n}=[\exp\left(hf_c/k_BT_e\right)-1]^{-1}$, where $h$ is the plank's constant and $k_B$ is the Boltzmann's constant, the variance of Alice's quadrature $V_a = V_s+V_o$, and the variance of Eve's quadrature $V_e = 1$ \cite{2_ref_paper, neel_spd_channel_estimation_skr}. Furthermore, since this is a seminal work on the use of RIS for THz CV-QKD systems, we consider all the phases of the reflecting elements of the RIS to be set to the same angle, implying that $\phi_k = \phi, \forall k \in \left\{1,\ldots,K \right\}$. The plots are generated for three different cases owing to the practical consideration of a restricted quantum memory of Eve resulting in her storing one of the three ancilla modes $\left(\{e_{out_d},e_{qm}\}, \{e_{out_g},e_{qm}\},\{e_{out_f},e_{qm}\} \right)$ at any given time, which is employed to compute the quantum correlation with Bob's corresponding output quadrature.

Fig. \ref{f3} presents the plots of SKR versus the distance between Alice and Bob for varying MIMO configurations. It is to be noted that the values of $N_{T_X}$ and $N_{R_X}$ are taken to be the same leading to the value of $r$ (i.e. number of parallel channels) to be the same as the diversity branches. Moreover, the distances for the Alice-RIS pair and the RIS-Bob pair are considered to be $0.4$ and $0.7$ times the distance between Alice and Bob, respectively. It is observed that in the case when Eve computes the quantum correlation with Bob's output quadrature via the direct channel $\undb{H}_d$, the SKR increases at a given distance with an increase in the number of diversity branches. 
Moreover, the system can obtain a given value of SKR at a higher distance between Alice and Bob with an increase in the values of transmit and receive antennas. However, it is interesting to observe that both the values of the SKR and the distance increase when the measurements are carried out via the channel $\undb{H}_g$.
Moreover, these values increase even further for the measurements to be carried out over the channel $\undb{H}_f$. This implies that employing the RIS significantly improves the SKR of the system and can be used as a viable option even when the distance between the transmitter and the receiver increases. Specifically, this advantage of the use of RIS is prominent after the phase shifts have been employed to improve the strength of the signal which corresponds to the measurement in the channel $\undb{H}_f$. Thus, RIS is a viable technology to improve the performance of a system for the transmission of secret keys between Alice and Bob. The following plots study the effect of the RIS in terms of the number of reflecting elements and the choice of the phase shits on the performance enhancement of the system under consideration. The comparison plots with similar system parameters without considering RIS are also presented in Fig. \ref{f3}. It is observed that the SKR values considering RIS are superior as compared to the SKR values obtained for the non-RIS system. Furthermore, a comparison of the SKR values computed using the analysis in \cite{natural_comm2017pirandola} is also presented. It is observed that the SKR values obtained from \cite{natural_comm2017pirandola} outperform our system only for the case when Eve stores the ancilla modes $\left\{ e_{out_d} , e_{qm} \right\}$. This is due to the fact that the effect of RIS is prominent for the other two cases of the storage of ancilla modes by Eve. 

\begin{figure*}
     \centering
     \begin{subfigure}[b]{0.334\textwidth}
         \centering
        \includegraphics[width=\textwidth,height=9cm]{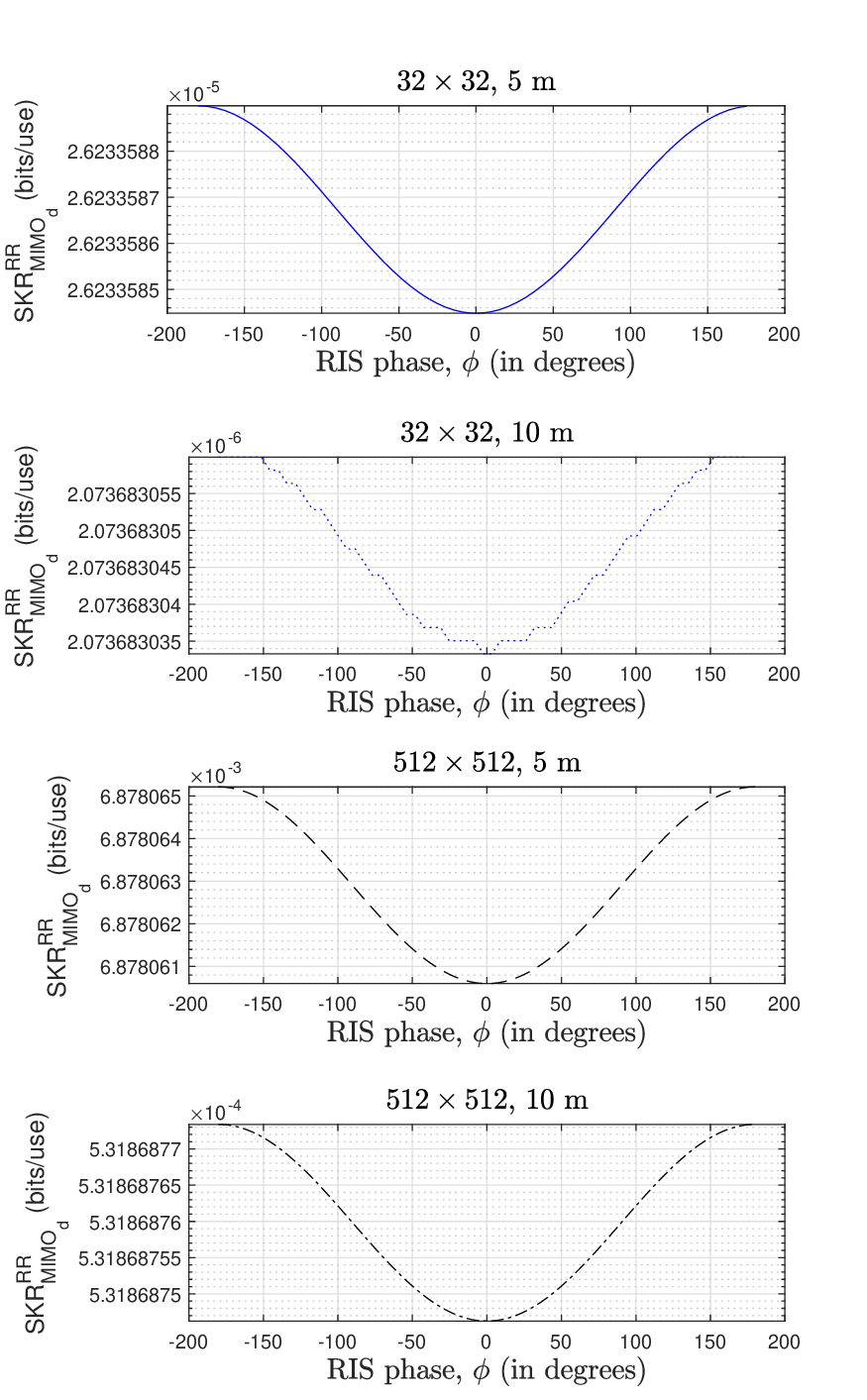}
         \caption{Ancillary mode $\left\{ e_{out_d} , e_{qm} \right\}$}
         \label{fig;5a}
     \end{subfigure}
     \hspace{-0.3cm}
     \begin{subfigure}[b]{0.334\textwidth}
         \centering
        \includegraphics[width=\textwidth,height=9cm]{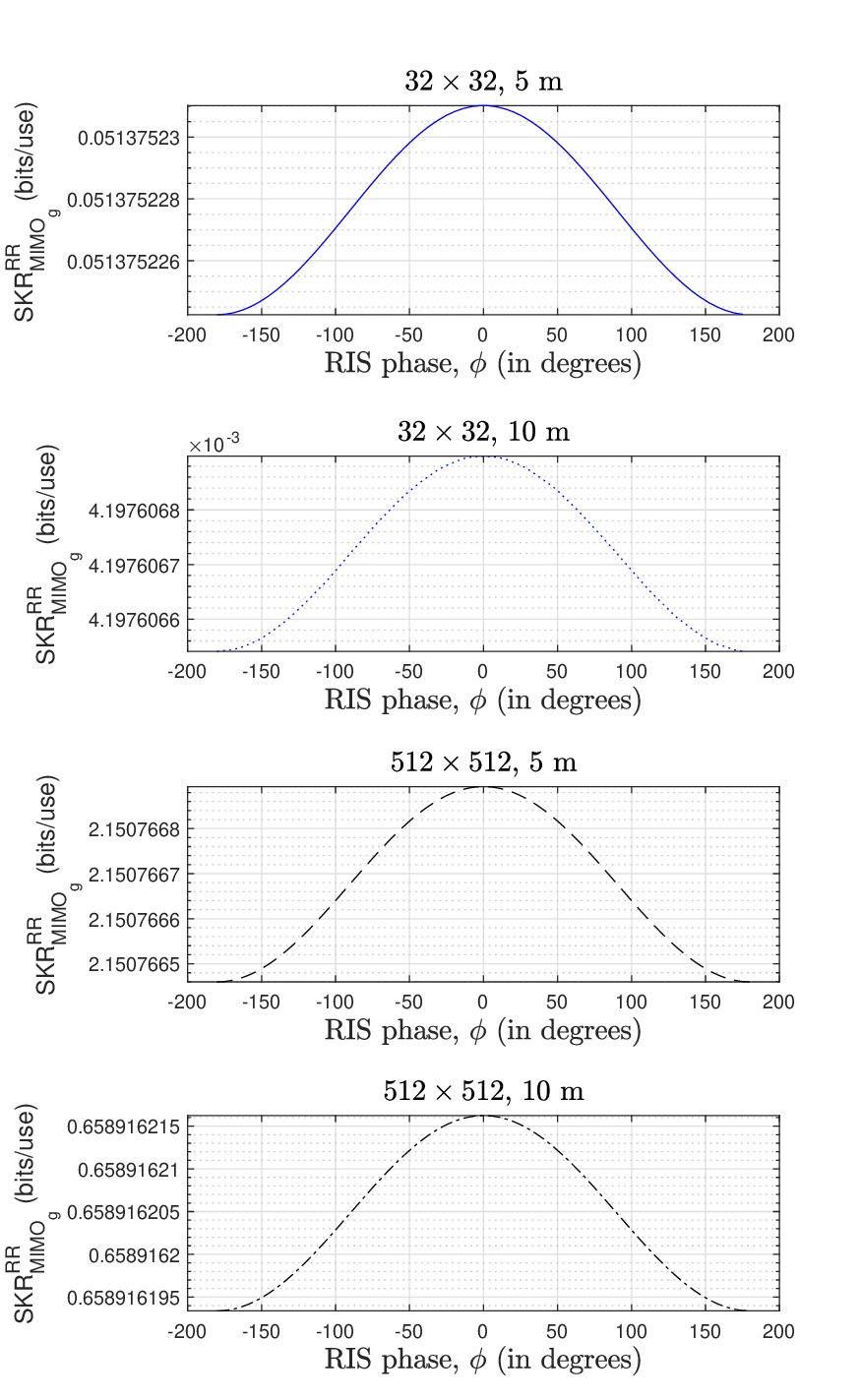}
         \caption{Ancillary mode $\left\{ e_{out_g} , e_{qm} \right\}$}
         \label{fig;5b}
     \end{subfigure}
     \hspace{-0.3cm}
     \begin{subfigure}[b]{0.334\textwidth}
         \centering
        \includegraphics[width=\textwidth,height=9cm]{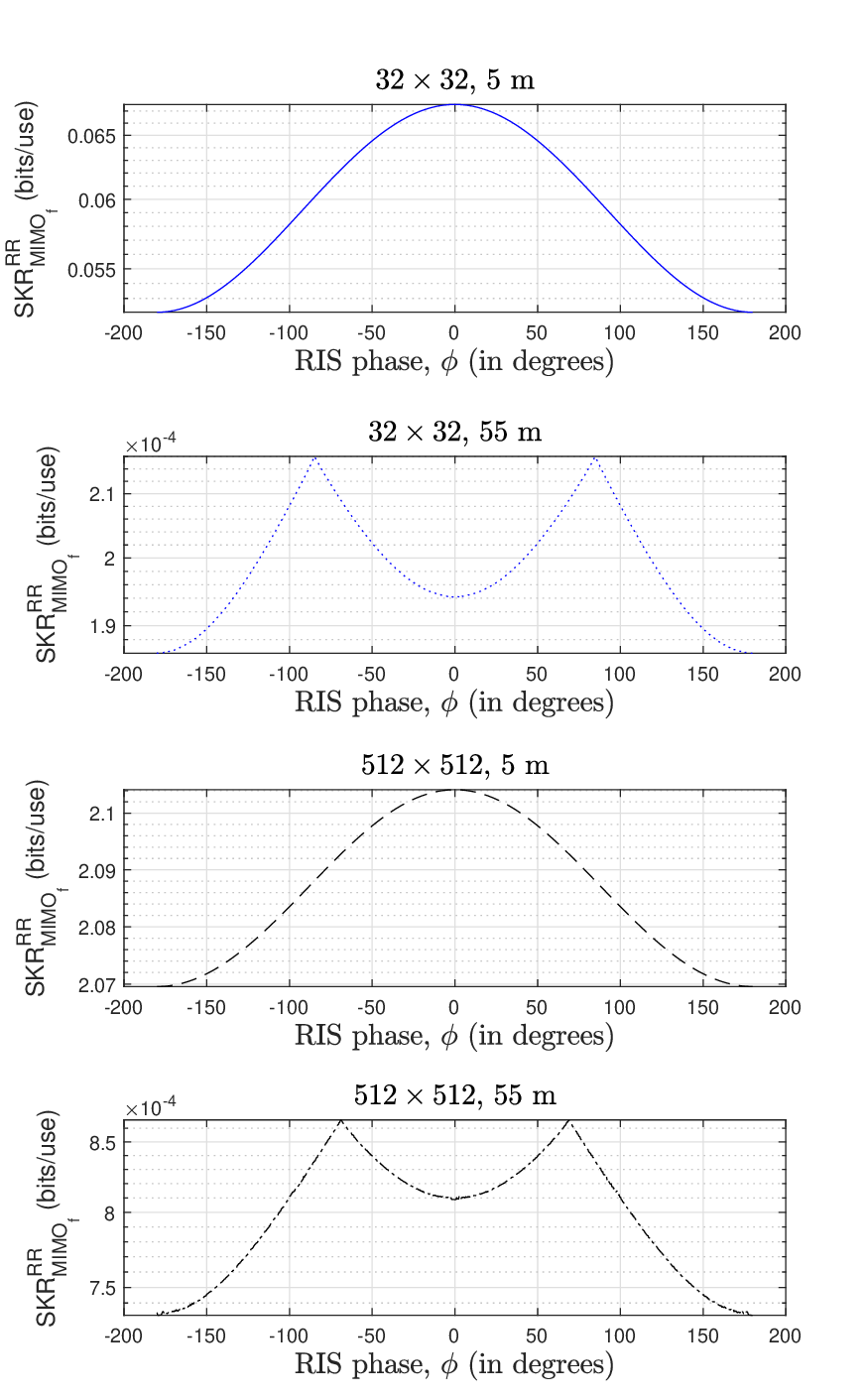}
         \caption{Ancillary mode $\left\{ e_{out_f} , e_{qm} \right\}$}
         \label{fig;5c}
     \end{subfigure}
        \caption{SKR versus RIS phase shifts $\phi$ for $K=100$, $N_{T_x}=N_{R_X}=32, 512$, $f_c$ = 10 THz, $d_a=0.5\lambda_c$ and for Eve storing the ancilla modes (a) $\{e_{out_d},e_{qm}\}$ with distance between Alice and Bob as $5,10$ m, (b) $\{e_{out_g},e_{qm}\}$ with distance between Alice and Bob as $5,10$ m, and (c) $\{e_{out_f},e_{qm}\}$ with distance between Alice and Bob as $5,55$ m.}
        \label{f5}
        \vspace{-0.3cm}
\end{figure*}
The plots for the SKR versus the distance between Alice and Bob for varying numbers of RIS elements are presented in Fig. \ref{f4}. Similar to Fig. \ref{f3}, the distances for the Alice-RIS pair and the RIS-Bob pair are considered to be $0.4$ and $0.7$ times the distance between Alice and Bob, respectively. In general, it is observed that an increase in the number of reflecting elements improves the SKR of the system. However, from Fig. \ref{fig;4a} it is observed that the effect of the RIS on the performance improvement is distinguishable only for lower distances between Alice and Bob, implying that increasing $K$ does not play any role in the increase in the SKR of the system for the case when Eve computes Bob's quadrature via the direct channel $\undb{H}_d$ for a higher distance between Alice and Bob. A similar trend is observed in the case of Eve obtaining the measurements via $\undb{H}_g$ as observed in Fig. \ref{fig;4b}. However, the advantage of increasing the number of RIS elements to improve the SKR of the system is seen in Fig. \ref{fig;4c}, implying that when Eve measures the quantum correlation via the channel between RIS and Bob, an increase in the number of reflecting elements would yield a higher SKR value even with increasing distance between Alice and Bob.

\begin{figure}[t]
    \centering
    \includegraphics[width= 8cm, height=5cm]{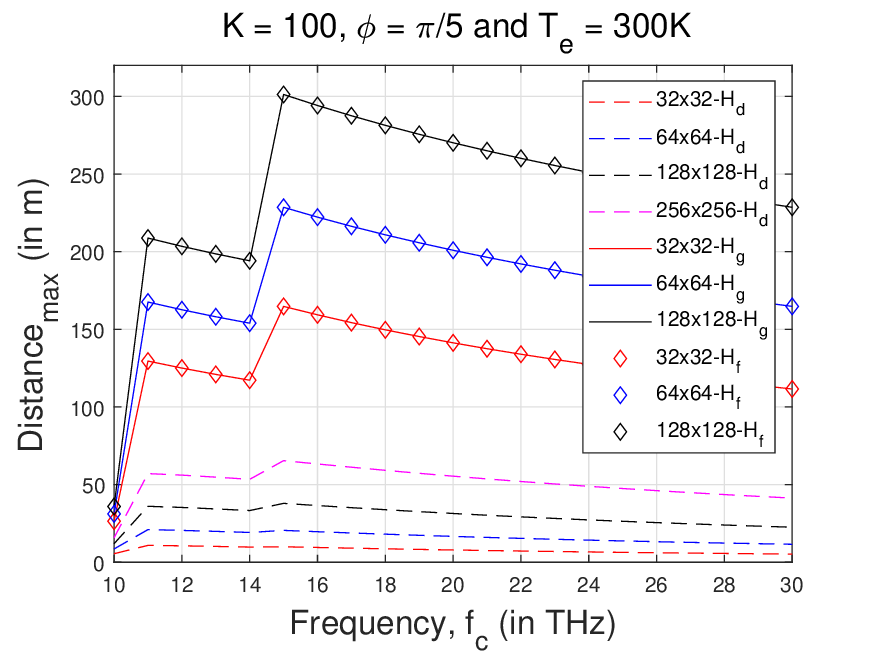}
    \caption{Distance between Alice and Bob versus the operating carrier frequency for $K=100$, $\phi=\pi/4$, $N_{R_X}=N_{T_X}=32,64,128$, and the cases when Eve stores the ancilla modes from the corresponding channel.}
    \label{f6}
    \vspace{-0.5cm}
\end{figure}
\begin{table}[t]
\caption{Optimal phase shifts of RIS elements for $N_{T_X}=N_{R_X}=16,32,64,128,256,512,1024$ and distance between Alice and Bob as $50,80$ m}
\begin{center}
\begin{tabular}{|c|p{2.5cm}|p{2.5cm}|}
    \hline
    \multirow{2}{*}{$N_{T_X} \left(= N_{R_X}\right)$}&\multicolumn{2}{c|}{Optimal phase shift, $\phi$, of RIS elements}\\ \cline{2-3}
    {}&Distance between Alice and Bob $=50$ m & Distance between Alice and Bob $=80$ m \\ \hline
    16 & \hspace{0.9cm}$85.85^{\circ}$ & \hspace{0.9cm} $89.29^{\circ}$ \\ \hline
    32 & \hspace{0.9cm}$83.56^{\circ}$ & \hspace{0.9cm} $89.29 ^{\circ}$\\ \hline
    64 & \hspace{0.9cm}$83.56^{\circ}$ & \hspace{0.9cm} $83.56^{\circ}$\\ \hline
    128 & \hspace{0.9cm}$77.83^{\circ}$ & \hspace{0.9cm} $83.56^{\circ}$\\ \hline
    256 & \hspace{0.9cm}$72.10^{\circ}$ & \hspace{0.9cm} $83.56^{\circ}$ \\ \hline
    512 & \hspace{0.9cm}$66.37^{\circ}$ & \hspace{0.9cm} $77.83^{\circ}$\\ \hline
    1024 & \hspace{0.9cm}$54.91^{\circ}$ &\hspace{0.9cm} $77.83^{\circ}$ \\ \hline
\end{tabular}
\end{center}
\label{tab1}
\vspace{-0.8cm}
\end{table}
Fig. \ref{f5} depicts the plots of variation of the SKR with the RIS phase shift for varying MIMO configuration and distance between Alice and Bob. It is interesting to observe that in Fig. \ref{fig;5a}, i.e., for the case of Eve correlating with Bob's quadrature via the channel $\undb{H}_d$, the choice of $\phi=0$ would yield the least SKR and the choice of $\phi = \pi$ would result in the highest value of SKR. This implies that the reflecting elements in the RIS should shift the phase of the incident signal by $\pi$ to maximize the SKR of the system. On the contrary, as observed in Fig. \ref{fig;5b}, the SKR of the system is maximized when the values of the phase shifts are zero. This implies that to maximize the SKR of the system, the RIS should not shift the phase of the incident signal when Eve attempts to decrypt the secret keys via measurements on the channel $\undb{H}_g$. It is to be noted that these trends in Figs. \ref{fig;5a} and \ref{fig;5b} are not dependent on the change in the number of diversity branches or distance between Alice and Bob. However, the dependency of the SKR on the phase shifts of the RIS elements for the case of Eve measuring Bob's quadrature via $\undb{H}_f$ is dependent on these system parameters. It is observed that for a lower distance between Alice and Bob, the optimal phase shifts should be equal to zero. As the distance between Alice and Bob increases, the SKR at $\phi=0$ reduces and does not remain at the maximum anymore. It is observed that there exists a value of the phase shift at which the SKR of the system is the maximum and this trend becomes more prominent with the increase in the distance between Alice and Bob and for lower configurations of the MIMO system. These optimal phase shift values for varying system parameters are tabulated in Table \ref{tab1}.

It is observed from all the previous plots that, apart from the SKR, the distance between Alice and Bob is a crucial parameter to be increased to improve the coverage of the system under consideration. Thus, the variation of the distance with the operating carrier frequency for varying MIMO configurations is presented in Fig. \ref{f6}. It is observed that the maximum distance increases with the increase in the diversity of branches. Moreover, although there is not a direct trend between the distance and the carrier frequency, it is observed that the effect of the RIS to improve the distance is significant for the case when Eve measures the ancilla modes corresponding to the channel $\undb{H}_f$. Thus, the use of RIS is most practical to improve the secrecy and coverage of the considered system when the RIS is placed close to the transmitter.

\section{Conclusion}
This paper considered a RIS-assisted MIMO CV-QKD wireless communication system where the transmitter, Alice, transmits secret keys to Bob via a direct channel between them and by the wireless channels assisted by a RIS employing $K$ reflecting elements. The eavesdropper, Eve, attempted to intercept the encrypted signal by attacking all the channels and computing the quantum correlation of its ancilla modes with the output obtained by Bob employing homodyne measurement. Considering a practical consideration of a limited quantum memory of Eve, she was considered to store the correlation from one of three channels, namely the direct channel $\undb{H}_d$ between Alice and Bob, the channel $\undb{H}_g$ between Alice and the RIS, and the channel $\undb{H}_f$ between RIS and Bob. Furthermore, transmit and receive beamsplitters were employed based on SVD to convert the $N_{R_X} \times N_{T_X}$ THz MIMO system into $r$ parallel SISO channels. For this system model under consideration, novel closed-form expressions for the SKR were derived for all three cases of Eve's measurements. Numerical results were presented to corroborate the analytical framework and to study the dependency of the various parameters on the performance of the system. It was observed that RIS plays a significant role in increasing the SKR and the transmission distance of the system. The advantage of utilizing RIS was found to be specifically significant when Eve measured the ancilla mode obtained for the channel between the RIS and Bob. Moreover, it was observed that the phase shifts of the RIS play a major role in maximizing the SKR of the system, with the optimal phase shifts to be a) $\pi$ for Eve measuring the ancilla modes corresponding to $\undb{H}_d$, b) zero for Eve measuring the ancilla modes corresponding to $\undb{H}_g$, and c) varying optimal phase shifts for Eve measuring the ancilla modes corresponding to $\undb{H}_f$ at lower and higher distances between Alice and Bob. More practical considerations of the phase shifts of all the RIS elements to be different, the effect of atmospheric impairments, and imperfect channel state information pave the future research directions of this work.
\bibliographystyle{IEEEtran}
\bibliography{IEEEabrv,bibliography}
\end{document}